\documentclass[a4paper,fleqn]{cas-sc}

\usepackage{setspace}
\usepackage{caption}
\usepackage{subcaption}
\usepackage{listings}
\usepackage{xcolor}
\usepackage{longtable}
\usepackage{todonotes}
\usepackage{seqsplit}

\usepackage[authoryear]{natbib}
\setcitestyle{authoryear,open={(},close={)},citesep={;}}

\newcommand{\longstring}[1]{{\ttfamily\seqsplit{#1}}}

\def\tsc#1{\csdef{#1}{\textsc{\lowercase{#1}}\xspace}}
\tsc{WGM}
\tsc{QE}
\tsc{EP}
\tsc{PMS}
\tsc{BEC}
\tsc{DE}

\newboolean{showcomments}
\setboolean{showcomments}{false} 

\ifthenelse{\boolean{showcomments}}
  {\newcommand{\nb}[2]{
  \fbox{\bfseries\sffamily\scriptsize#1}
     {\sf\small$\blacktriangleright$\textit{\textcolor{red}{#2}}$\blacktriangleleft$}
   }
  }
  {\newcommand{\nb}[2]{}
   
  }

\begin{document}
\let\WriteBookmarks\relax
\def\floatpagepagefraction{1}
\def\textpagefraction{.001}
\shorttitle{You Shall not Repackage! Demystifying Anti-Repackaging on Android}
\shortauthors{A. Merlo, A. Ruggia, L. Sciolla, L. Verderame}

\title [mode = title]{You Shall not Repackage! Demystifying Anti-Repackaging on Android}


\author[1]{Alessio Merlo}[orcid=0000-0002-2272-2376]
\cormark[1]
\ead{alessio@dibris.unige.it}

\address[1]{DIBRIS - University of Genoa, Via Dodecaneso, 35, I-16146, Genoa, Italy.}

\author[1]{Antonio Ruggia}\ead{antonio.ruggia@dibris.unige.it}

\author[1]{Luigi Sciolla}\ead{luigi.sciolla@dibris.unige.it}


\author[1]
{Luca Verderame}[orcid=0000-0001-7155-7429]
\ead{luca.verderame@dibris.unige.it}

\cortext[cor1]{Corresponding author}


\maketitle
\begin{abstract}
App repackaging refers to the practice of customizing an existing mobile app and redistributing it in the wild. In this way, the attacker aims to force some mobile users to install the repackaged (likely malicious) app instead of the original one. This phenomenon strongly affects Android, where apps are available on public stores, and the only requirement for an app to execute properly is to be digitally signed.

Anti-repackaging techniques try counteracting this attack by adding logical controls in the app at compile-time. Such controls activate in case of repackaging and lead the repackaged app to fail at runtime. On the other side, the attacker must detect and bypass the controls to repackage safely. 
The high-availability of working repackaged apps in the Android ecosystem suggests that the attacker's side is winning. 

In this respect, this paper aims to bring out the main issues of the current anti-repackaging approaches.
The contribution of the paper is three-fold: 1) analyze the weaknesses of the current state-of-the-art anti-repackaging schemes (i.e., Self-Protection through Dex Encryption, AppIS, SSN, SDC, BombDroid, and NRP), 2) summarize the main attack vectors to anti-repackaging techniques composing those schemes, and 3) show how such attack vectors allow circumventing the current proposals. The paper will also show a full-fledged attack to NRP, the only publicly-available anti-repackaging tool to date. 

\end{abstract}

\section{Introduction}
\label{sec:intro}

During the last five years, Android strengthened its leadership among mobile operating systems with more than 75\% of the market share on average~\citep{statcounter:2019}.
Most of this success is due to the app development process, which  is easier compared to other mobile operating systems: in fact, Google currently offers a plethora of platforms that support the developers in designing, implementing, testing, and sharing their app ~\citep{android-developer-guide}.
As a consequence, the number of Android apps in the Google Play Store reached 2.8M in 2020 \citep{appbrain_android_apps}.

From a security standpoint, Android apps are the main target for attackers, as they allow to reach a large number of mobile users. In fact,
since a mobile app is available in public app markets (e.g., Google Play Store and Samsung App Store), an attacker can easily retrieve the app bundle (i.e., the $apk$ file, which is basically an archive containing the app). Then, the attacker can reverse engineer the $apk$ \citep{2011-android-reversing}, inject some malicious code, and re-distribute a modified version of the original app. This kind of attack is called \textsl{repackaging}.

Android repackaging is a serious problem that may affect any unprotected app. A repackaged app can cause money losses for developers as the attacker can re-distribute paid apps for free, remove/redirect ads earnings, or make the app willingly unusable to negatively impact the reputation of the developer.
Furthermore, past studies already demonstrated that the 86\% of Android malware is contained in repackaged apps \citep{10.1109/SP.2012.16} and 77\% of the top 50 free apps
available in Google Play have been plagiarized by malicious actors and distributed in alternative app stores \citep{free-repacked}. This is indeed a promising strategy for malware developers, as embedding the malicious code in a repackaged version of a popular and well-ranked app speeds up its spread.

It is worth pointing out that repackaging is intrinsically related to the Android development life-cycle, where an $apk$ just needs to have a valid signature in order to be successfully installed and execute properly, without requiring any guarantee on the actual identity of the signer (i.e., the developer does not need a valid public key certificate issued by a trusted certificate authority). Therefore, an attacker can modify an existing $apk$ and then repackage and sign it with its own self-generated private key. 

In order to make the repackaging phase more challenging for attackers, in recent years several techniques have been proposed in literature. Such techniques can be divided into two sets, namely \textsl{repackaging detection} and \textsl{anti-repackaging}: the first set aims at recognizing repackaged apps, while the latter one focuses on inserting proper controls in the app, such that a repackaged version of the same app would not work properly. 

While repackaging detection techniques have been widely discussed and analyzed in literature \citep{Zhan2019ACS, 10.1109/TSE.2019.2901679}, no systematic analysis of anti-repackaging schemes is still available. To this aim, in this paper, we provide an extensive analysis of such techniques. The contribution of the paper is three-fold: 
after introducing some background and the threat model related to app repackaging in Section \ref{sec:background}, we discuss the state of the art of anti-repackaging schemes in Section \ref{sec:sota}. Then, in Section \ref{sec:attacking_model} we summarize the attacking techniques that can be used to circumvent anti-repackaging. In Section \ref{sec:attack} we will show how existing anti-repackaging schemes can be circumvented to produce a fully-working repackaged version of a protected app. 
We also discuss a full-fledged attack to the most recent anti-repackaging technique to date, which is the only one whose source code is publicly available. We show how to bypass the protections and successfully repackage a real app (i.e., Antiminer \citep{antimine-android}).
Finally, in Section \ref{sec:conclusion} we draw some conclusions, pointing out some guidelines to improve the robustness and the reliability of the next-generation anti-repackaging techniques. 



\section{App Repackaging: Backgroud and Threat Model}\label{sec:background}

In this section, we briefly recap the basics of app development and distribution in Android. Then, we discuss a threat model for app repackaging, and we summarize the main categories of repackaging detection and anti-repackaging techniques.

\subsection{Android App life-cycle}\label{subsec:android_lifecycle}

\paragraph{App development.} An Android app is usually developed in Java or Kotlin and leverages XML files to represent the graphical user interface of the app. From now on, we will explicitly refer to Java code, although the same techniques are valid also in case of Kotlin-based apps. There are other ways to develop Android apps, for example, hybrid apps. Hybrid apps are mainly built using web technologies and run over a WebView (e.g., Apache Cordoba \citep{cordova}), and are likewise weak against repackaging because the same techniques used by attackers allow exploiting also JavaScript components.

Each Android app is distributed and installed as an Android Package ($apk$) file. Until 2018, developers had to build their $apk$ file to distribute their app.
Since 2018 Google provides a new publishing format, i.e., the Android Application Bundle (AAB) \citep{android-app-boundle}, which allows building optimized $apk$ for each device configuration;
 nonetheless, the app is delivered to the final user as an $apk$\footnote{Therefore, we will refer to $apk$ only, hereafter.}. 

In a nutshell, an $apk$ file is a zip archive containing all the necessary files to run the first execution of the app.
Each $apk$ is signed with its developer's private key and also contains the corresponding public certificate of the developer. This mechanism grants the integrity of the $apk$, but cannot provide any authentication, as the developer certificate does not need to be issued by a trusted certificate authority. Therefore, Android performs only basic verification on the $apk$ structure and its integrity, thereby leaving the decision to install the app (and, consequently, to trust the source of the app) to the final user. 

Regarding app repackaging, the most relevant files and folders contained in the $apk$ are:
\begin{itemize}
  \item \emph{META-INF/}: it contains the developer's public certificate and the app signature;
  \item \emph{lib/}: it contains the native libraries for each specific architecture (e.g., x86, armeabi\-v7a). An app could execute native C/C++ code through the Java Native Interface (JNI) \citep{android_ndk};
  \item \emph{res/}: it contains the binary resources, such as images or hybrid apps content;
  \item \emph{AndroidManifest.xml}: it describes the structure of the app and its components. In this file the developer defines app permissions, entry points, intent filters, and more;
  \item \emph{classes.dex}: it contains the compiled Java/Kotlin code in the form of bytecode, i.e., a high level representation of the machine code, that could be converted into Smali code, i.e., a human readable format, with off-the-shelf tools like Backsmali \citep{backsmali}. 
\end{itemize}

\paragraph{Application distribution.}
Android apps are mainly delivered through proper software repositories, called \textsl{app stores}, which allow developers to publish their $apk$. iOS relies on a single, centralized market  (i.e., the Apple Store), which is directly controlled by Apple. On the contrary, a distinctive feature of Android is the possibility to distribute the same app on dozens of different app stores beyond the official one (i.e., the Google Play Store \citep{google_play_store}).  The most popular app stores differ from region to region. In Europe and US, the Google Play Store has the greatest market share \citep{android-markets-2019}. However, in 2019, in China, \textsl{Tencent My App} was the biggest Android app store with up to 25.5\% of the market share, followed by the \textsl{360 Mobile Assistant} and \textsl{Xiaomi App Store}; in this country, the Google Play Store is just tenth place \citep{china-markets, android-markets-2019}. 

Whenever a user downloads an app from any store, Android carries out a soft check on the app certificate before installation, with the final decision left to the user~\citep{android-publish-app}: we recall that an $apk$ has to be signed by the developer, but Android continues the installation even if the signer of the developer's certificate is unknown~\citep{android-apk-authenticity}.
Since the only constraint is a valid signature, it is also possible to distribute the $apk$ directly (e.g., through email or urls) without the need to upload the app on an app store. 

\subsection{App Repackaging: Threat model}\label{sec:threat-model}

\begin{figure*}[!ht]
  \includegraphics[width=\textwidth]{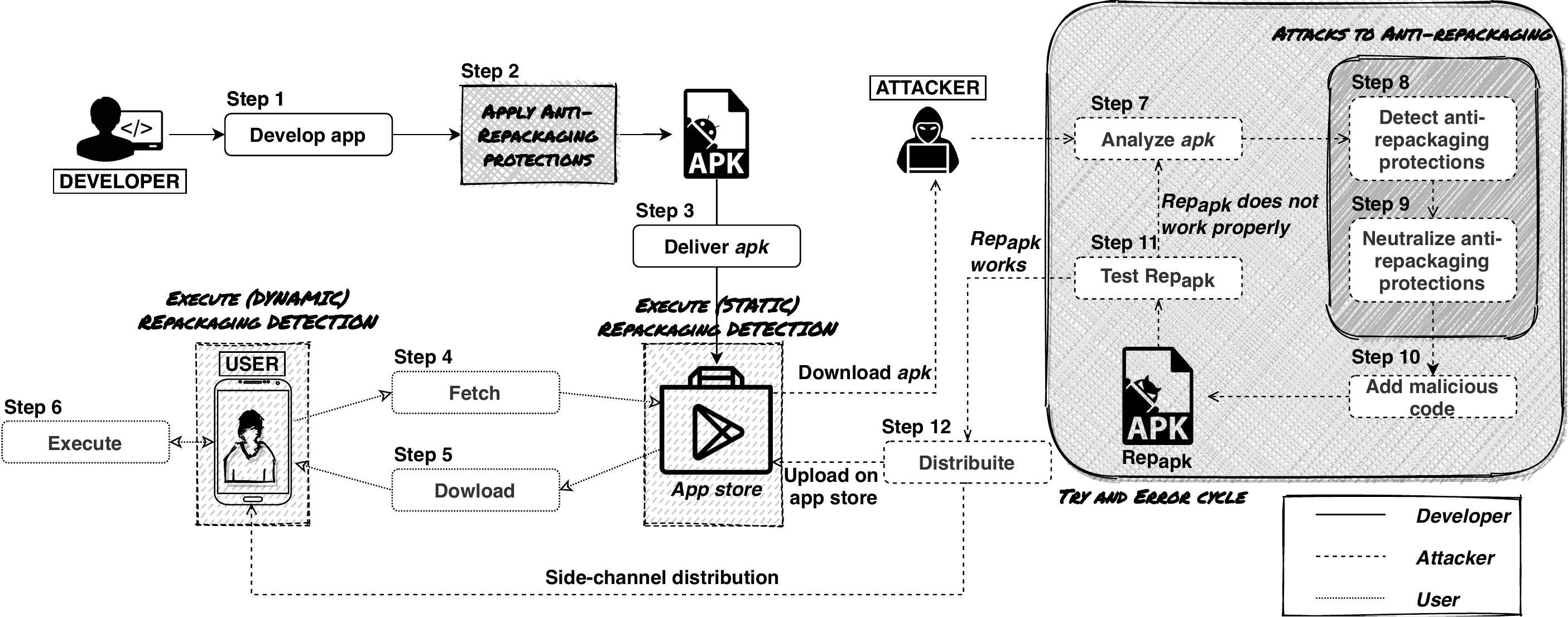}
  \caption{App repackaging Threat model.}
  \label{fig:threat-model-design}
\end{figure*}

In this section, we discuss a threat model for app repackaging. The model is depicted in Fig. \ref{fig:threat-model-design}. 

The threat model involves three actors, namely the Developer, the Attacker, and the User. The Developer implements the app (Step 1), and may also add some anti-repackaging protection in the code before building the $apk$ (Step 2), in order to counteract repackaging attempts. In general, applying an anti-repackaging technique means to add some extra code that may lead, for instance, the repackaged app to crash. Then, the Developer builds and signs the $apk$ and delivers it to the app store (Step 3). 

In an ideal (i.e., attack-free) scenario, the User fetches the store for apps (Step 4), then chooses and downloads the original $apk$. The validity of the signature and the integrity of the $apk$ is verified automatically by Android, by leveraging the developer's public key certificate provided in the $apk$. Then, the $apk$ is installed successfully and the User can execute the app (Step 6).

In an actual use case, an attacker can download the developer's $apk$ by behaving as any normal user. Then, the Attacker decompiles and analyzes the $apk$ (Step 7), and tries to detect the anti-repackaging protections (Step 8). The Attacker can decode the $apk$ using Apktool~\citep{apktool} and perform static analysis on the smali source code. Most of the smali code could be further decompiled back into Java code~\citep{10.1007/3-540-45937-5_10}. If static analysis is not sufficient, an attacker can dynamically infer the app behavior at runtime, by relying, for instance, on Frida \citep{frida}. A fruitful combination of static and dynamic analysis can allow the Attacker to detect the anti-repackaging protections added by the developer in the app code.
In case some protections are detected, the Attacker tries to neutralize them (Step 9) by deactivating the corresponding extra code, accordingly. Then, he can add some malicious code (Step 10) and build and signs a repackaged version of the $apk$ (i.e., $Rep_{apk}$) with a valid self-generated public key certificate. Finally, the Attacker tests $Rep_{apk}$ on its own devices (Step 11). In case it works, the Attacker can try distributing $Rep_{apk}$ (Step 12) both on official channels (i.e., app stores) and side-channels (e.g., by email, on website or through phishing attacks). Otherwise, the 
Attacker needs to carry out further analysis on the $apk$, and repeat Steps 7 to 11 (also known as the \textsl{try and error cycle}) until she obtains a working $Rep_{apk}$. 

The attack succeeds whether some users install $Rep_{apk}$ instead of the original $apk$. This situation is rather common (see Section \ref{subsec:android_lifecycle}), as during the installation process Android just executes some soft checks on the certificate used to sign the $apk$, but it does not verify the developer's identity. As a consequence, this provides no hints to the final user on the actual reliability of the $apk$. 

The execution of $Rep_{apk}$ leads the user to execute the malicious code on her own device. The malicious code may have different aims, like e.g., access the app premium features for free or redirect revenue to cause financial loss to the developer, gain access to the user's private in app data (e.g., credentials), or make the app willingly unstable to affect the reputation of the original app.

\subsection{Repackaging Detection vs. Anti-Repackaging}\label{sec:threat-mitigation}

This section points out the main features of both \textsl{repackaging detection} and \textsl{anti-repackaging} techniques. 

\subsubsection{Repackaging Detection}\label{subsec:repackaging-detection}
Repackaging detection focuses on recognizing repackaged apps in app stores or on user's devices, with the aim to limit the spread of repackaged app (i.e., Step 12 in Figure \ref{fig:threat-model-design}).
As a lot of Android malware is contained in repackaged apps, detecting a repackaging can allow identifying malware. Due to this, several repackaging detection techniques have been proposed in recent years. These techniques can be divided into static and dynamic approaches \citep{10.1016/j.procs.2016.02.006}, and further grouped into five different categories \citep{10.1109/TSE.2019.2901679}. 
\begin{itemize}
    \item \textbf{Static (or offline techniques)} are adopted on \textsl{app stores} to identify uploads of repackaged apps which are clones of original apps. 
    Such techniques strongly relies on \textsl{symptom discovery} or \textsl{app similarity}. Symptoms \citep{10.1109/TSE.2019.2901679} are tracks, such as strings, left by the repackaging process itself. Similarity checks usually involve a two-step process in which each app is first profiled according to a set of distinctive features; then, apps with almost identical feature profiles are identified.  Such techniques make strong use of machine learning, both supervised \citep{0.1109/SPW.2016.33, Lin2013IdentifyingAM} and unsupervised~\citep{10.1007/978-3-319-23829-6_30}. 
    Static analysis needs a considerable amount of resources both in order to generate the decision model and to make the final decision and can be efficiently applied only on the server side.    
    \item \textbf{Runtime (or online/dynamic techniques)} are directly executed on the user's device to evaluate the app.
    These approaches extract specific information both at install-time and at run-time. For instance, the technique discussed in \citep{10.1145/2484313.2484315} puts a watermark inside the app which is checked by a third authority at runtime.
\end{itemize}

Both static and runtime techniques rely on an external trusted authority responsible for the detection or repackaged apps and the enforcement of countermeasures. While in the first case the external authority can prevent the spreading of repackaged apps by blocking their publication, in the latter case the detection may occur only after the $apk$ has been installed on the user's device; as a consequence, it may be too late for preventing the malware activation. On the other hand, since users could install repacked apps from external sources, offline techniques could help, but are insufficient to fully prevent the spreading of repackaged apps.

\paragraph{Attacking Repackaging Detection.} An attacker successfully circumvents a repackaging detection technique once it is able to build repackaged apps that go undetected both on the app store and on the devices. According to the characteristics of the current detection techniques, this means that the repackaged app must appear very different from the original app (to bypass similarity checks), hide repackaging identifiers (to avoid symptom discovery), and behave as the original app to circumvent dynamic analysis.  
It is also worth pointing out that the granularity of the Android ecosystem - which allows to deliver apps through an undefined number of app stores and side channels (e.g., apps available at some URIs or delivered by email) - makes hard to build up a reliable, pervasive and full-fledged repackaging detection deployment. In fact, this would require to deploy static analysis mechanisms on each source of apps (i.e., app stores and servers), as well as dynamic analysis solutions on each mobile device.

\subsubsection{Anti-Repackaging}\label{subsec:repackaging-avoidance}
Anti-Repackaging - also known as \textsl{repackaging avoidance} or \textsl{self-protection}
 - aims to protect an app from being successfully repackaged. As opposed to repackaging detection, anti-repackaging is applied by the developer to the app code before building the $apk$ (Step 2) and aims to make any repackaged version of the same app crash. On the bad side, the attacker aims to detect and neutralize the anti-repackaging protection after reversing the $apk$ (Steps 8 and 9). 
From a technical standpoint, anti-repackaging techniques insert some logic controls, called \textsl{detection nodes} or \textsl{anti-tampering}, inside the $apk$ . These nodes detect when an $apk$ or a system tampering has been carried out.
Since the first aim of the attacker is to recognize anti-repackaging, the requirements of anti-tampering checks are i) to go undetected and hide inside the app code, and ii) in case of detection, to avoid being de-activated. To achieve such results, anti-tampering checks may rely on code obfuscation, code virtualization, and anti-debugging techniques.  We briefly introduce anti-tampering checks and such techniques.

\paragraph{Anti-tampering checks.} In a nutshell, anti-tampering checks are self-protecting functions which aim to detect any modification on the original app. There exist several methods to detect tampering in mobile apps, as highlighted in \cite{BERLATO2020102463}. 

The most relevant anti-tampering techniques are:
\begin{itemize}
    \item \textbf{Signature checking}: the most trivial control, which checks the certificate contained in the $apk$. This approach could detect any modification applied to the original $apk$;
    \item \textbf{Code integrity}: it checks whether some specific part of the code has been tampered with, by computing its signature at run-time. As an example, the first bytes of the \emph{.dex} file contains the hash of its content. This method detects tampering only if the attacker has modified the checked methods;
    \item \textbf{Resource integrity}: it checks the signature of some resources at run-time. 
    This technique differs from the previous ones as it checks tampering attempts in the app resources (e.g., images and other binary files) instead of the code;
    \item \textbf{Installer verification}: it checks if the installing app comes from trusted app stores. This technique is applied under the assumption that tampered apps are more likely to be distributed on unofficial app stores.
\end{itemize}
Anti-tampering checks dynamically react to any modification either in the app or in the executing environment. 
Once some checks are detected by the attacker (Step 8 in  Figure \ref{fig:threat-model-design}), the same attacker tries to bypass them by following the \textsl{try and error} approach, i.e., she modifies the app (in order to try removing checks), then she repackages and tests the app to check whether it works properly. If the attempt fails, then the attacker may try to search for other checks and/or modify the app in a different way.

\paragraph{Code obfuscation.} Obfuscation approaches modify the bytecode or the source code of an app, without changing its behavior, in order to make some manual or automatic analysis more difficult. There exist a lot of off-the-shelf obfuscation tools both open source (e.g., \cite{aonzo2020obfuscapk}) and commercial (e.g.,\cite{proguard}). 

\paragraph{Code virtualization.}
Code virtualization is a specific obfuscation scheme. In this approach, instructions are mapped to semantically-equivalent virtual instructions. A virtual machine (VM) executes inside the app and interprets the virtual instructions. Two different approaches are proposed in \cite{10.1145/2557547.2557558, 2019-exploiting-binary}; the first one works at DEX level, while the latter focuses on native libraries. 
Similar to obfuscation, this approach aims at complicating the reverse engineering process of an app, but it does not introduce any anti-tampering check.

\paragraph{Environment check.} 
Environment checks aim to detect the reliability of the environment (e.g., the OS) in which apps are installed. These checks include anti-debugging controls which aim to detect and avoid dynamic analysis checks (e.g., debugger). Differently from passive obfuscation techniques, where the app code is syntactically modified to make it harder to understand, anti-debugging controls allow an app to actively react against malicious reverse engineering  at run-time \citep{BERLATO2020102463}.
The most widespread environment checks and anti-debugging techniques are:
\begin{itemize}
    \item \textbf{Emulator detection}: it checks whether an app is running on a real device or in an emulated environment;
    \item \textbf{Time check}: when an app executes with a debugger, its execution slows down. Time checks can be specific functions whose execution depend on the execution time: in case the execution time is higher than a given threshold, such functions take the app to an inconsistent state;
    \item \textbf{Check debuggable}: the \emph{AndroidManifest} contains a flag that enables the debugging mode: if this flag is set to false, an user is not able to attach any debugger to the app. The attacker can bypass this check by changing the manifest accordingly. Nonetheless, such modification could trigger some other anti-tampering checks.
\end{itemize}
Recently, more complex anti-debugging techniques have been put forward (e.g., ~\cite{10.1145/3015135.3015142, article}) to exploit proper features of the Android Runtime to hide anti-debugging instruction at runtime. 
Still, such an approach quickly became obsolete as it is compatible only with specific OS versions, i.e., up to Android API 25 (v.7.1)

\paragraph{Information hiding.} 
These techniques are commonly used by malware developers to modify the original code and hide some malicious payload. Similarly, in the anti-repackaging scenario, they are used to hide detection nodes and anti-tampering controls. 
Moreover, from the defending side, the most relevant information hiding technique is \emph{Code encryption}. This approach converts the program code into a ciphered string, which is decrypted at runtime. The decrypted code is executed via Java reflection.

\paragraph{Delayed execution.} This technique aims to delay the execution of some portion of code. In our scenario, \emph{delayed execution} can be applied to the code dealing with the repackaging detection response, in order to avoid the attacker to understand which detection node has been actually triggered.

\paragraph{Polymorphic code.} Polymorphic code mutates at each execution. This methodology is often used to hide malicious payload \citep{5975134} (from the malware side), or detection nodes (from the anti-repackaging side.

\paragraph{External Anti-tampering checks.} This category includes all the techniques relying on an external agent to execute anti-tampering checks. Examples are remote servers \cite{10.1145/2995306.2995315}, third-party $apks$ or OS kernel modules \cite{10.1007/978-3-030-05234-8_16}. 
One of the newest techniques in the state of the art is presented in \cite{10.1145/3371307.3371312}: here, the app gathers information about both its status and the device and the app itself, and send those data to an external server. The server is the authority which take the decision to continue the communication or label the app (or more in general the environment) as malicious. 
In order to achieve the goal of protection, these techniques face the challenge to create a communication channel between the app and the external authority. In addition, this channel itself has to be protected from repackaging, exposing it as a single point of failure.
\section{Anti-Repackaging: State of the art}\label{sec:sota}

\begin{table}[]
\begin{tabular}{|l|l|l|l|}
\hline
\multicolumn{1}{|c|}{\textbf{Anti-repackaging scheme}} &
  \multicolumn{1}{c|}{\textbf{Year}} &
  \multicolumn{1}{c|}{\textbf{Defending techniques}} &
  \multicolumn{1}{c|}{\textbf{Tampering checks}} \\ \hline
Self-Protection through dex Encryption &
  2015 &
  \begin{tabular}[c]{@{}l@{}}- Code encryption\\ - Code obfuscation\end{tabular} &
  - Code integrity \\ \hline
Stochastic Stealthy Networks (SSN) &
  2016 &
  \begin{tabular}[c]{@{}l@{}}- Detection nodes\\ - Delayed response\\ - Java reflection\end{tabular} &
  - Signature checking \\ \hline
Anti-repackaging Immune System (AppIS) &
  2017 &
  \begin{tabular}[c]{@{}l@{}}- Polymorphic code\\ - Detection nodes\\ - Relying on native code\end{tabular} &
  \begin{tabular}[c]{@{}l@{}}- Code integrity\\ - Resource integrity\end{tabular} \\ \hline
Self-Defending Code (SDC) &
  2018 &
  - Code encryption &
  - Code integrity \\ \hline
BombDroid &
  2018 &
  \begin{tabular}[c]{@{}l@{}}- Detection nodes\\ - Code encryption\end{tabular} &
  \begin{tabular}[c]{@{}l@{}}- Code integrity\\ - Signature checking\\ - Resource integrity\end{tabular} \\ \hline
Native Repackaging Protection (NRP) &
  2019 &
  \begin{tabular}[c]{@{}l@{}}- Detection nodes\\ - Code encryption\\ - Relying on native code\end{tabular} &
  - Code integrity \\ \hline
\end{tabular}
\caption{Anti-repackaging schemes: defending techniques and anti-tampering checks.}
\label{table:protection_schema_recap}
\end{table}


This section presents the state of the art of anti-repackaging schemes on Android, ordered chronologically. For each proposal, we will describe the idea, the protection scheme applied on the app before release, the runtime protection during the app execution, as well as the experimental results. Furthermore, in Table \ref{table:protection_schema_recap} we summarize the defending techniques and anti-tampering checks leveraged by the anti-repackaging schemes. 


It is worth pointing out that our analysis does not include protection schemes that rely on \emph{external anti-tampering checks}.
This choice is due to the fact that the use of external trusted parties, either remote (e.g., a verification server) or local (e.g., a controller app), increases the protection mechanisms' attack surface and, consequently, the range of attacks available.
For instance, protection schemes leveraging remote entities may suffer from network-based vulnerabilities \citep{7299933, 8068748} that could interfere with the protection mechanisms. Moreover, app developers cannot assume the presence of third-party solutions installed on the device. For such reasons, we focused only on protection schemes that allow an app to protect itself without relying on any external third-party.

\subsection{Self-Protection through \texttt{dex} Encryption} 
\label{subsec:dyn-self-prot-encrypt-2015}
\cite{7299906} proposed an anti-repackaging scheme aimed at complicating both the reverse engineering and the repackaging process. The main idea is to encrypt the \texttt{classes.dex} file in the $apk$, and dynamically decrypt and execute it at runtime. 

\begin{figure}[!ht]
  \includegraphics[width=\textwidth]{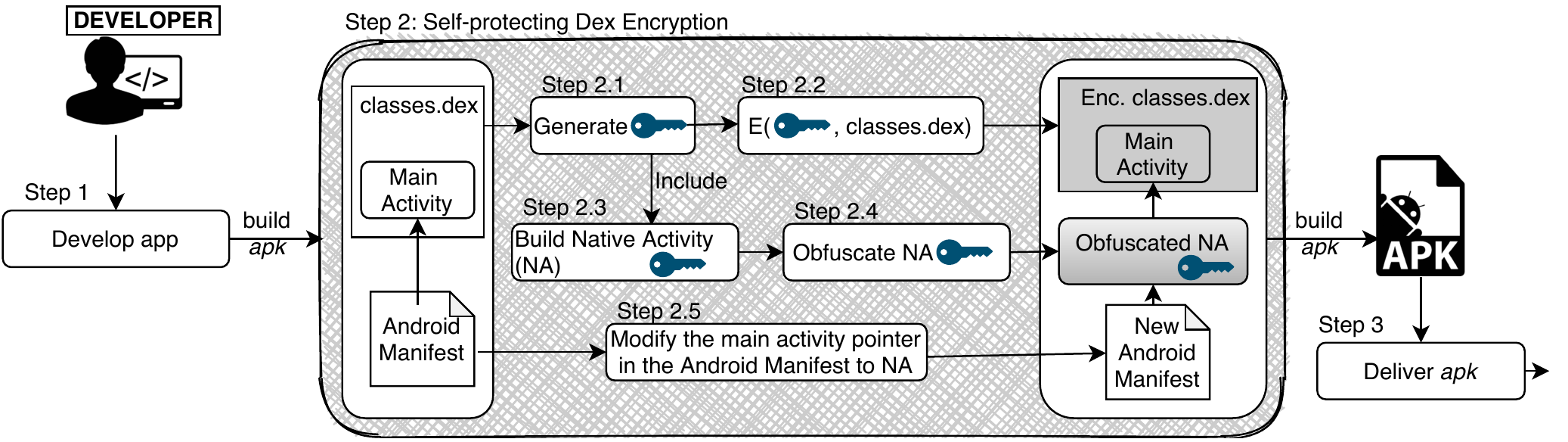}
  \caption{The Self-Protection through dex Encryption protection scheme.}
  \label{fig:dynamic-scheme}
\end{figure}

\paragraph{Protection scheme:}

The approach, depicted in Fig. \ref{fig:dynamic-scheme}, applies to the compiled $apk$, that is modified according to the following steps: 
\begin{enumerate}
    \item A XOR key is generated (Step 2.1) and applied to encrypt the whole \texttt{classes.dex} file(s) (Step 2.2);
    \item An native activity\footnote{An activity is an app window containing the UI.} is automatically built (Step 2.3) and compiled in a shared library. Such activity is called native, as it is written in native code (C/C++ code) instead of Java/Kotlin code as the other activities. The XOR key is added to the native activity;
    \item The native activity is obfuscated using the OLLVM tool \citep{ieeespro2015-JunodRWM} (Step 2.4);
    \item The Android Manifest is configured to execute the native activity (written in C/C++) rather than the original main activity\footnote{The Main Activity is the first activity prompted to the user when the app is executed.} when the app starts executing (Step 2.5).
\end{enumerate}
Finally, the $apk$ file is generated, signed and delivered (Step 3).
    
\begin{figure}[!ht]
  \includegraphics[width=\textwidth]{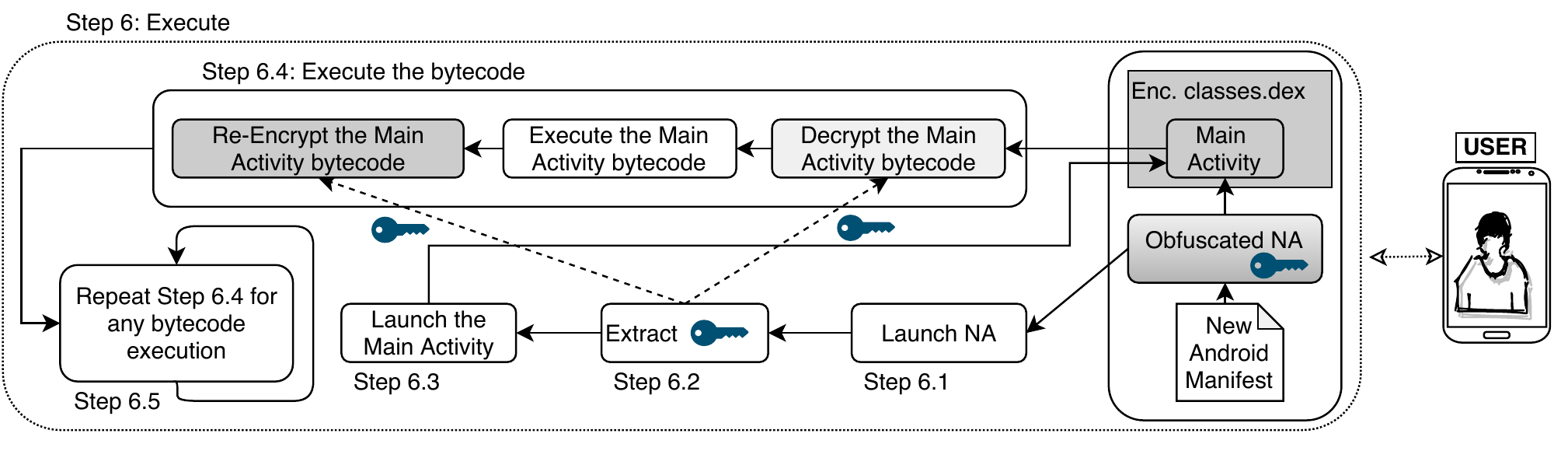}
  \caption{The Dynamic Self-Protection and Tamperproofing behavior at runtime.}
  \label{fig:dynamic-scheme-runtime}
\end{figure}

\paragraph{Runtime behavior:}
When the app executes, the protection scheme behaves as depicted in Fig. \ref{fig:dynamic-scheme-runtime}. At first, the modified manifest forces the execution of the native activity (Step 6.1). The native activity begins by extracting the XOR key (Step 6.2) and launches the main activity (step 6.3). As the execution of the main activity requires access to the plain bytecode, the XOR key is applied to decrypt only the part of the code required for the proper execution of the main activity. Then, after the execution is completed, the same bytecode is re-encrypted using the same key (Step 6.4). The approach is time-sliced, thereby decrypting parts of the \texttt{classes.dex} file on-demand, i.e., maintaining the bytecode unencrypted only when it needs to execute (Step 6.4).  

\paragraph{Evaluation:}
The adoption of XOR-based encryption of the bytecode is aimed at implementing a tamper-proof checksum.  
This work has been evaluated in \cite{7299906} on 749 apps from the F-Droid app store \citep{f-droid}. Due to implementation and process choices, only 312 apps can be properly protected. For instance, it was not possible to protect apps which make use of Java reflection. The experimental setup leveraged several Android devices equipped with Android 4.4.2, and stimulated through the Android Monkey UI Exerciser framework \citep{monkey} at runtime. Clearly, the adoption of encrypting/decrypting cycles, as well as the addition of extra code, negatively impacts the app time and space complexity. In fact, the experiments indicate a mean execution overhead of 440\% and an increased $apk$ size by 183\% on average, as reported in the original work. 

In \cite{7299906}, the reliability of the proposed protection scheme is evaluated according to four metrics: method exposure (both absolute and percent value), and instructions exposure (both absolute and percent value). A method or an instruction is \textsl{exposed} when it is decrypted. Experimental results show wide variance. On average, about 83\% of methods - which amount to the 80\% of the bytecode instructions - are exposed. 

\subsection{Stochastic Stealthy Networks} 
\label{subsec:ssn-2016}
\cite{7579771} put forward another self-protecting scheme against repackaging, named Stochastic Stealthy Network (SSN). 

\begin{figure}[!ht]
  \includegraphics[width=\textwidth]{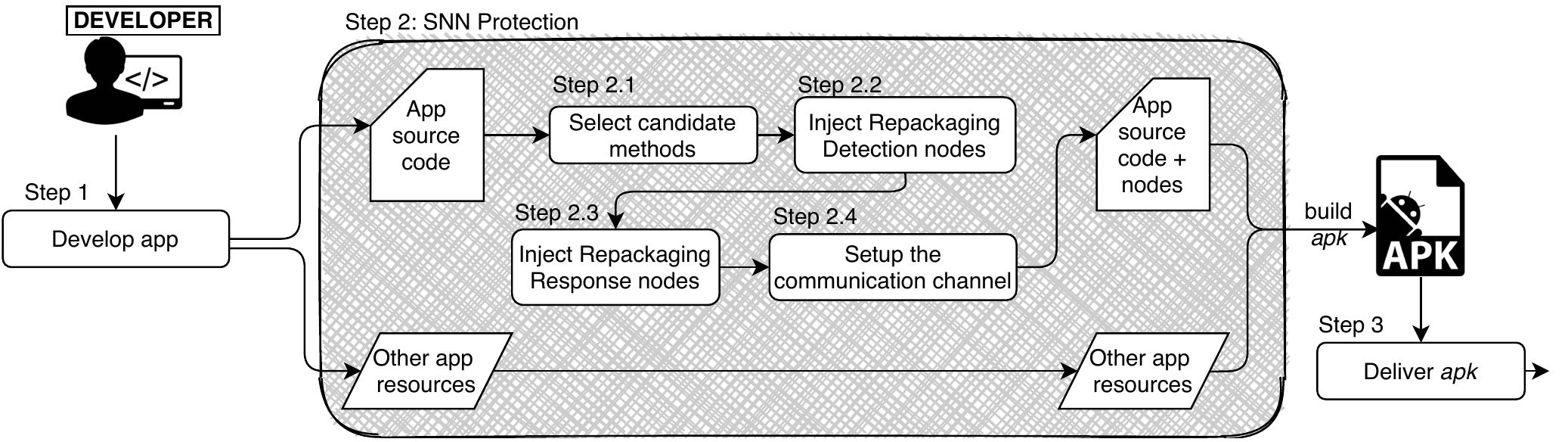}
  \caption{The Stochastic Stealthy Network (SSN) protection scheme.}
  \label{fig:ssn-scheme}
\end{figure}

\paragraph{Protection scheme:}
The idea is to distribute (and hide) a set of guards (i.e., \textsl{repackaging detection} nodes) into the app source code. When executed, such guards trigger proper decision points (i.e., \textsl{repackaging response} nodes) that in case tampering is detected, make the app crash. 
More in detail, the protection workflow, depicted in Figure \ref{fig:ssn-scheme}, begins by selecting a set of candidate methods in the Java code, according to some heuristics (Step 2.1). Then, a set of repackaging detection nodes are injected in such methods (Step 2.2), as well as the set of repackaging response nodes (Step 2.3). Such latter nodes take decisions according to a stochastic function. Moreover, a communication channel is added in the code to allow both kinds of node to communicate at runtime (Step 2.4). The communication channel is made by extra variables and methods added in the app code. Finally, SSN builds, signs, and delivers an $apk$ with code extended with the repackaging nodes.

It is worth noticing that the selection of candidate methods must take into consideration the overhead due to the execution of detection and response nodes: to this aim, good candidate methods must be seldom invoked at runtime. Furthermore, the detection nodes must be strongly distributed over the set of candidate methods, to increase stealthiness. 

\begin{figure}[!htb]
  \includegraphics[width=\textwidth]{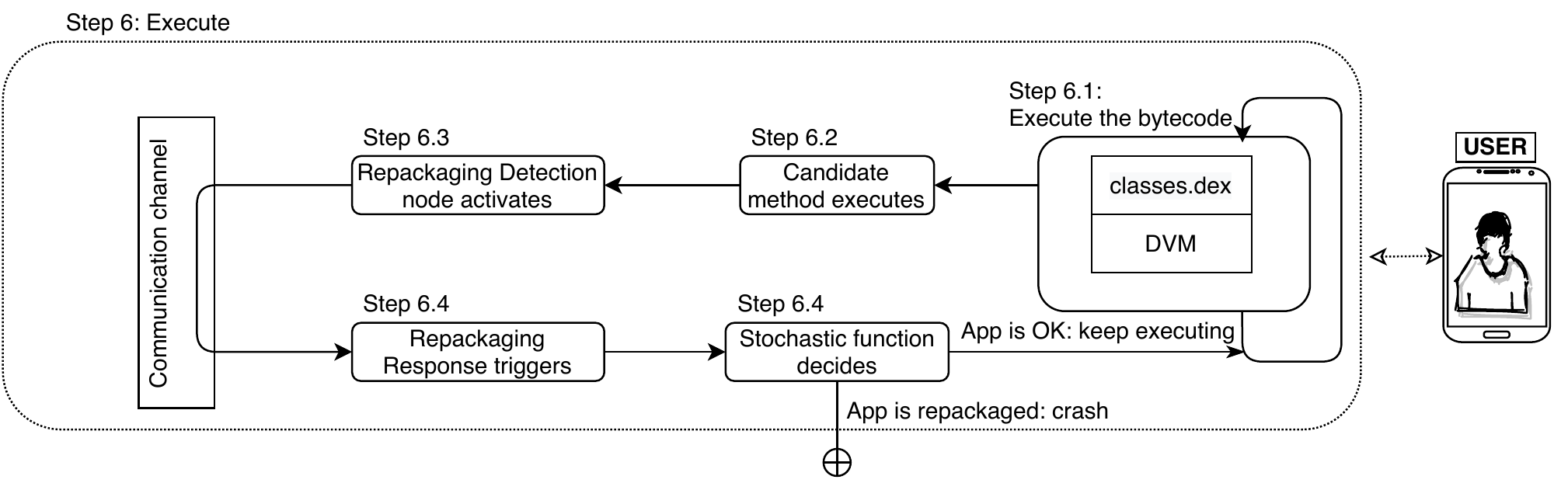}
  \caption{The SSN behavior at runtime.}
  \label{fig:ssn-scheme-runtime}
\end{figure}

\paragraph{Runtime behavior:}
Fig. \ref{fig:ssn-scheme-runtime} depicts the SSN behavior during app execution. Whenever a candidate method executes (Step 6.2) and activates a repackaging detection node (Step 6.3), then the same node connects to the corresponding response node(s) through the communication channel. If the stochastic function recognizes the method as repackaged (i.e., modified) then some execution is triggered. Such execution may lead to modify specific variable values or raise specific exceptions\footnote{Unfortunately, the paper does not provide such details.} that can result in program crashes or delayed logical malfunctions. 

\paragraph{Evaluation:}

The SSN approach only verifies the integrity of candidate methods by relying on the developer's public key verification. However, SSN supports several other predefined templates. For instance, reflection is used to call functions such as $getPublicKey()$ and \longstring{generateCertificate()}. 

The viability of the SSN approach has been empirically evaluated in the original work \citep{7579771} on a set of 600 apps, belonging to ten app categories in F-Droid. Experiments were conducted on emulators equipped with Android 4.1.
The average time overhead is between 6.4\% and 12.4\%, which is far lower than the time overhead of the previous approach, discussed in Section \ref{subsec:dyn-self-prot-encrypt-2015}. Furthermore, each protected app works properly, and no space overhead is noticed. 

\subsection{AppIS: Protect Android Apps Against Runtime Repackaging Attacks}\label{subsec:appis-2017}
\cite{8368344} proposed an app reinforcing framework, named AppIS. The idea of AppIS is to add security units as guards with an interlocking relationship between each other, in order to build redundant and reliable anti-repackaging checks.

\begin{figure}
  \includegraphics[width=\textwidth]{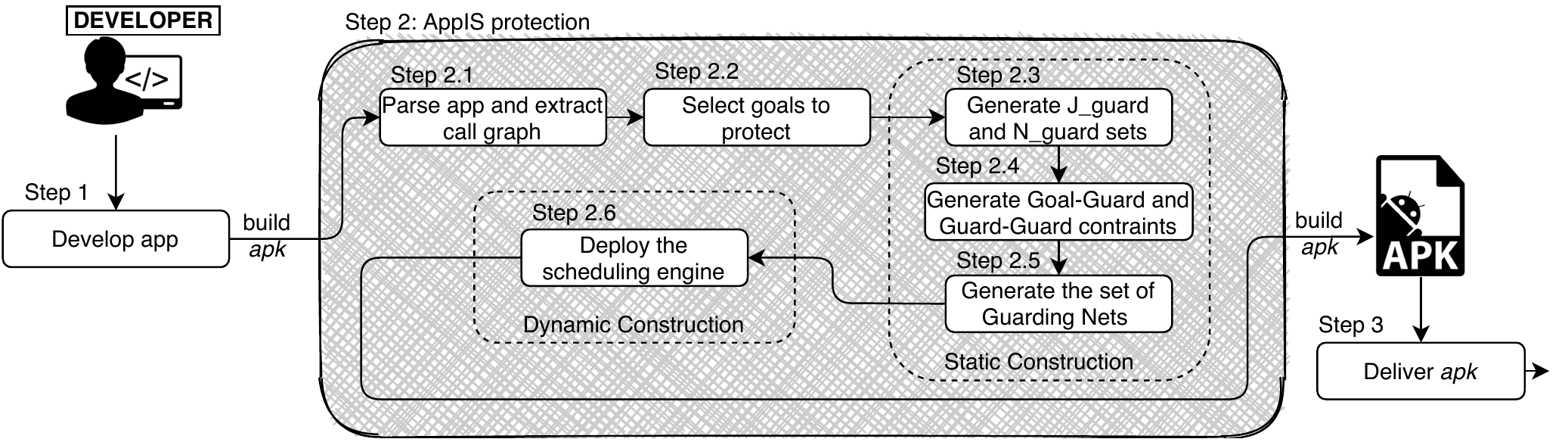}
  \caption{The AppIS protection scheme.}
  \label{fig:appis-scheme}
\end{figure}

\paragraph{Protection scheme:}
Fig. \ref{fig:appis-scheme} depicts the protection workflow of AppIS. The AppIS workflow begins by parsing the compiled $apk$ to extract its corresponding function call graph (Step 2.1), leveraging static analysis techniques at the state of the art. From such graph, AppIS selects a set of sensitive methods, resources, or data to protect (Step 2.2), named \textsl{goals}.  Then, AppIS generates a set of guards aimed to protect goals (Step 2.3). Guards can be hosted both in the bytecode (i.e., \longstring{J\_Guard}) and the native code (i.e., \longstring{N\_Guard}). Each goal must be controlled at least by a guard (\longstring{Guard-Goal}) and each guard must be likewise controlled by at least two other guards (\longstring{Guard-Guard}) to improve redundancy and reliability to attacks. According to previous rules, AppIS generates a set of Guard-Goal and Guard-Guard constraints (Step 2.4), which are taken into consideration to build a set of \textsl{guarding nets} among goals and guards, that satisfies all constraints. Moreover, such set is used to construct a scheduling engine that at runtime chooses which guarding net to activate, according to the triggered goal and the execution status. Finally, AppIS builds and delivers the protected $apk$, containing goals, guarding nets, and the scheduling engine.

\paragraph{Runtime Behavior:} At runtime the scheduler generates a guarding net (Step 6.1). In this way, a different guarding net is generated at each execution, in order to make the app resilient against cumulative attacks (i.e., multiple executions of the app in a sandboxed environment to detect guards). 
During the execution (Step 6.2), whenever a guard is triggered (Step 6.3), the corresponding guards check the integrity of the goal (i.e., \longstring{J\_Guard}) or the guard (i.e., \longstring{N\_Guard}) (Step 6.4) by comparing the checksum of the goal calculated during the static construction (i.e., Steps 2.3 to 2.5) with the checksum obtained at runtime on the same goal (Step 6.4). If the two checksums do not match, each guard detects a repackaging, and the app is directly terminated.

\begin{figure}[!ht]
  \includegraphics[width=\textwidth]{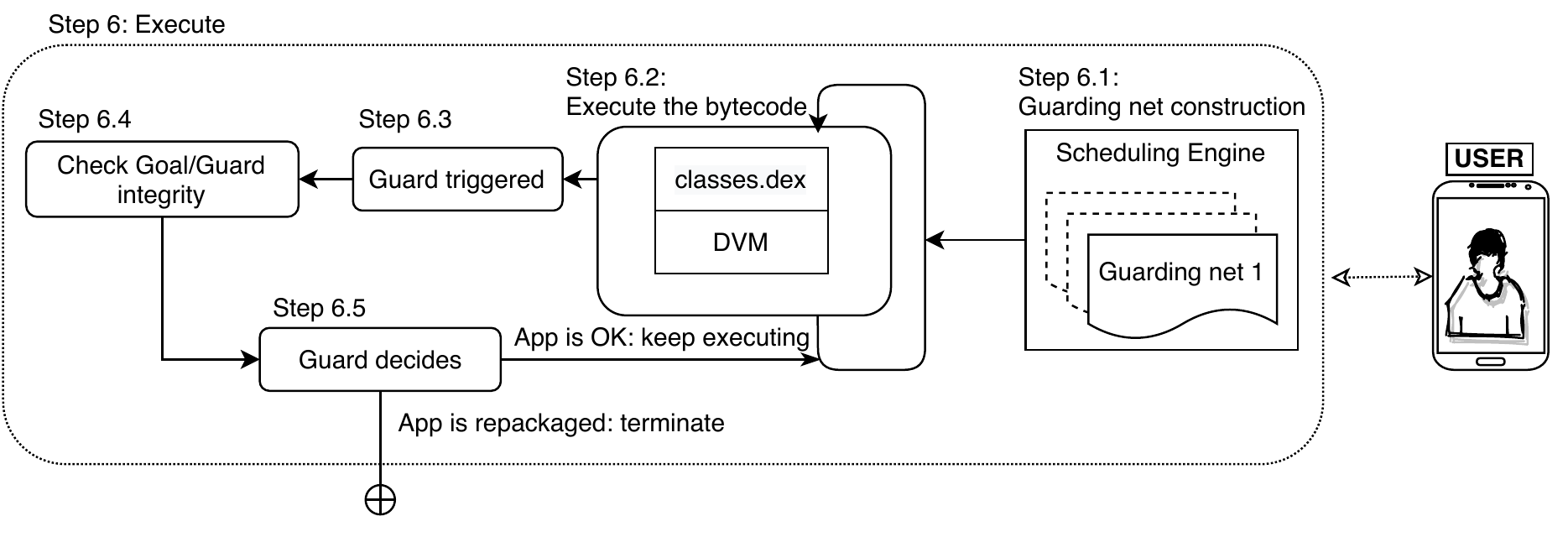}
  \caption{The AppIS behavior at runtime.}
  \label{fig:appis-runtime}
\end{figure}

\paragraph{Evaluation:}
AppIS improves the previous proposals by adding networks of guards and a non-deterministic behavior at runtime, that complicates the attacker analysis activity. In \cite{8368344}, AppIS has been evaluated on a very reduced set of only 8 apps, collected from different sources, on almost outdated Android versions (spanning from 4.4 to 6). Experimental results indicate that no app fails, and overheads are reasonable w.r.t. the previous proposals; in fact,  the space overhead is up to 2\% in the worst case, while time overhead spans from 2\% to 140\%. However, we argue that the reduced number of apps is statistically insignificant.
The reliability of AppIS has been tested against two threats, namely the possibility for an attacker to i) obtain the collection of the guarding nets, and ii) to carry out cumulative attacks. According to the authors, in both cases, the AppIS approach revealed to be reasonably robust.  

\subsection{Self-Defending Code through Information Asymmetry}
\label{subsec:sdc-2018}
\cite{8186215} proposed a self-defending code (SDC) scheme, leveraging the information asymmetry between the developer and the attacker on the app code, i.e., it is reasonable to assume that the attacker has far less information than the developer on the app code, on average. Similarly to the approach discussed in Section \ref{subsec:dyn-self-prot-encrypt-2015}, the idea is to encrypt pieces of source code and decrypt (and properly execute) them at runtime only in case the app has not been repackaged. Unlike the previous work, each piece of code is encrypted with a different key.
The approach provides two schemes, where the first one requires the customization of the Android OS, and the second one can work on unmodified OSes. We will focus on the latter one, as we argue that the first scheme cannot scale in the wild, due to the customization needs. 
\paragraph{Protection scheme:}
\begin{figure}
  \includegraphics[width=\textwidth]{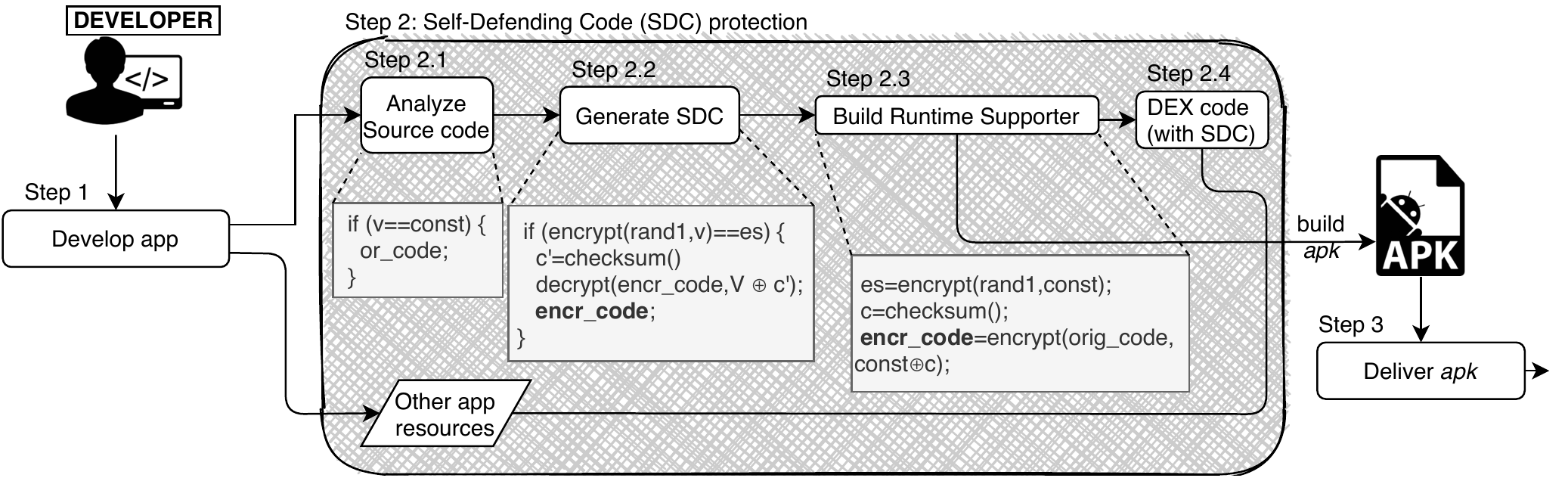}
  \caption{The SDC protection scheme.}
  \label{fig:sdc-scheme}
\end{figure}
The scheme (see Figure \ref{fig:sdc-scheme}) begins by analyzing the source code (Step 2.1) and selecting some qualified conditions (i.e., branches containing an equality check, where one of the operands is a constant value: i.e., \longstring{v==const}) in the code that needs to be protected. A qualified condition is then substituted by a self-defending code (SDC) (Step 2.2) that hosts the encrypted version of the original code contained in the body of the qualified condition. Once all the pieces of code have been encrypted, a component named \textsl{runtime supporter} is built. Such component contains all the encryption routines which allow understanding of how the modified code has been generated (Step 2.3); such routines are useful for reconstructing the original code at runtime.
It is worth pointing out that the encryption scheme is XOR-based and the key is related to the checksum of the original piece of code. Therefore, the original code can be decrypted correctly if and only if the app has not been modified. More in detail, w.r.t. to the sample code at Step 2.1 in Figure \ref{fig:sdc-scheme}, 
the information asymmetry between the developer and the attacker is the knowledge of the value \texttt{const}, removed from the source code at compiled time in Step 2.2, and the \longstring{encr\_code}. Furthermore, \longstring{encr\_code} is redundantly protected by a set of other - randomly selected -  SDC segments, some of which implement part of their functions as native code.  
Finally, the protected apk, composed by the DEX code extended with the SDC, the non-code resources, and the runtime supporter, is built and delivered.

\paragraph{Runtime Behavior:} At runtime, whenever an SDC is executed, the encrypted code is decrypted by leveraging the app checksum xored with the \texttt{v} value as shown in Step 2.2. If one of these differs from the original values, the decryption fails and the original code is not retrieved, thereby leading the app to crash.

\paragraph{Evaluation:}
During the evaluation phase of the authors \citep{8186215}, differently from previous proposals, SDC has been validated on a significant set of apps (i.e., 20,000) taken from Anzhi \citep{anzhi} and belonging to 15 categories. For each app, a set of more than 100 candidate branches has been selected and substituted with SDC segments. The testing phase involved both real users and automatic testing. In the first case, ten  SDC segments were triggered in an hour, while, in the latter case,  five segments were triggered within 24 hours. The time overhead is below 4\% for each app tested, while the space overhead is negligible.

\subsection{BombDroid: Decentralized Repackaging Detection through Logic Bombs} \label{subsec:bombdroid-2018}
\cite{10.1145/3168820} introduced the concept of \textsl{logic bombs} as anti-repackaging protections for Android apps. Such concept has been originally introduced in malware more than 20 years ago. The approach consists of hiding (i.e.,  cryptographically obfuscated) pieces of code that are executed once proper triggers (i.e., logical conditions) are activated. 
The authors implemented the approach in a tool called BombDroid. Like SDC (Section \ref{subsec:sdc-2018}), logic bombs are injected in qualified conditions.

\paragraph{Protection scheme:}
\begin{figure}[!ht]
  \includegraphics[width=\textwidth]{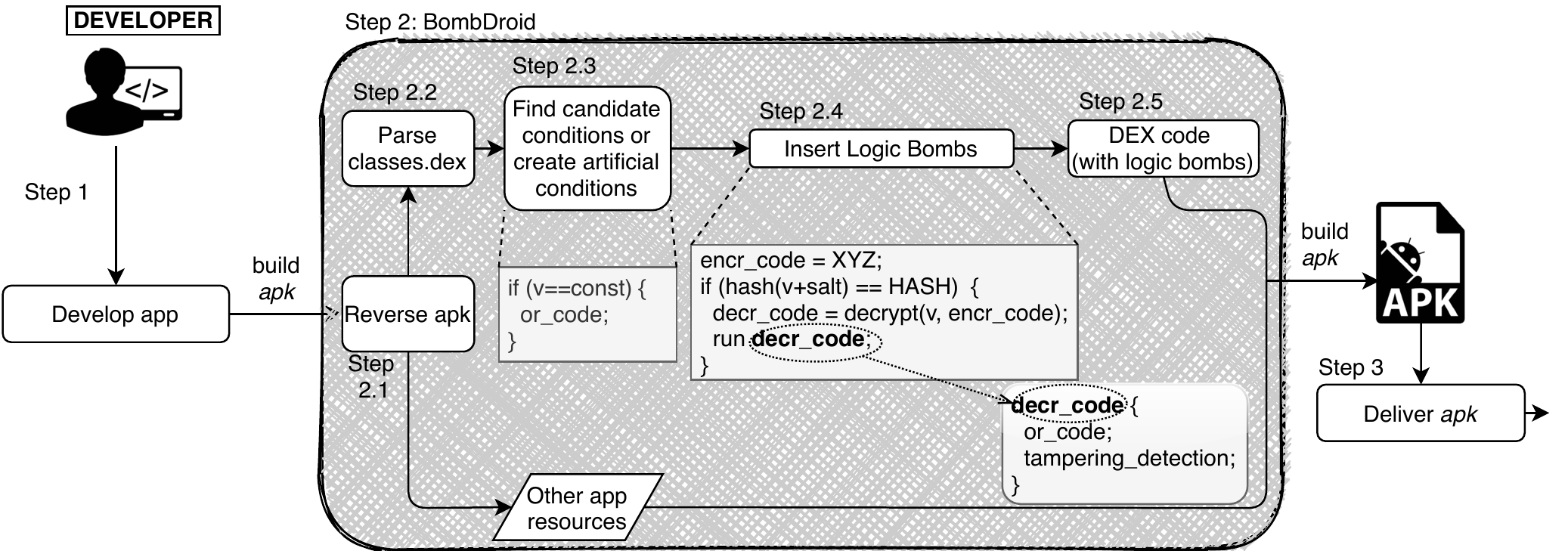}
  \caption{The BombDroid protection scheme.}
  \label{fig:bombdroid-scheme}
\end{figure}
The protection scheme of BombDroid (see Figure \ref{fig:bombdroid-scheme}) is very similar to the SDC one, and the idea is to selectively find some candidate condition in the app methods to inject logic bombs. The first difference is that BombDroid works directly on the $apk$ at the bytecode level, and not on the source code. The BombDroid protection scheme begins by reversing the $apk$ (Step 2.1), and parsing the \longstring{classes.dex} file (Step 2.2) through static analysis techniques to identify suitable methods (Step 2.3). In each selected methods, BombDroid identifies at least a qualified condition containing an equality between a variable and a constant value (\longstring{v==const}, see Figure \ref{fig:bombdroid-scheme}). The original code - contained in the corresponding block - is extended with some checks aimed at detecting tampering, and then, encrypted. The encrypted code will be  decrypted at runtime according to the hash value of the variable plus a salt, selected by BombDroid. The decryption key is the value of the constant \texttt{v}. The use of salt, instead of the checksum of the original method, is the main difference with SDC. The robustness of the scheme is likewise given by the robustness of the hash function, that does not allow guessing the value of \texttt{v}. Moreover, it is worth pointing out that the selected methods can have no qualified condition in principle: in this case, BombDroid can inject artificial qualified conditions to protect. Finally, BombDroid allows nesting logic bombs to improve stealthiness: for instance, with reference to the logic bomb in Step 2.4 of Figure \ref{fig:bombdroid-scheme}, another encrypted logic bomb can be inserted in the decrypted code between the original code and the tampering detection function. 

\paragraph{Runtime behavior:} At runtime, the logic bomb is executed, and the code is properly decrypted only if the value of the variable in the qualified condition is equal to \texttt{const} \footnote{This condition triggers the execution of \longstring{or\_code} in the original code of the candidate method.} and the logic bomb activates. 
Then, a tampering detection function is executed. BombDroid applies three anti-tampering techniques, i.e., at the app level, at the file level, and at the code snippet level. 
At compile time, BombDroid stores the original public key in the $apk$, as well as the values of some digests concerning resources and the \longstring{classes.dex} file, and some code snippets. At runtime, tampering functions calculate the digests of such resources and compare them with the stored one: tampering is detected whether one of such value differs from the stored one. In case tampering is detected, the user could be warned or the app could cause a negative user experience.

\paragraph{Evaluation:}
The authors assessed the reliability and the performance of BombDroid on a set of apps (i.e., 963) taken from F-Droid. They exploited Dynodroid \citep{10.1145/2491411.2491450} to execute each app for an hour, repeating the experiment 50 times per app. According to the authors' results \citep{10.1145/3168820}, the execution time overhead is almost negligible (i.e., <2.7\%) and the space overhead ranges from 8\% to 13\%. Furthermore, on average, the first logic bomb is triggered between 75 and 164 seconds after the app starts executing, and on average, only the 9.3\% of logic bombs are triggered. 

\subsection{Native Repackaging Protection}
\label{subsec:bombdroid-native-2019}
\cite{10.1007/978-3-030-45371-8_12} proposed an evolution of previous schemas. A prototype of this work is available on GitHub \citep{repackaging-protection}. Like BombDroid, this work is based on cryptographically obfuscated logic bombs, inserted in candidate methods with a qualified condition (i.e., \longstring{v==const}). However, the main difference is that integrity checks and a part of the original app code of the block are executed in native code. We refer to this scheme as \textsl{Native Repackaging Protection}\footnote{Authors do no provide a name for this scheme.} (NRP). 

\begin{figure}
  \includegraphics[width=\textwidth]{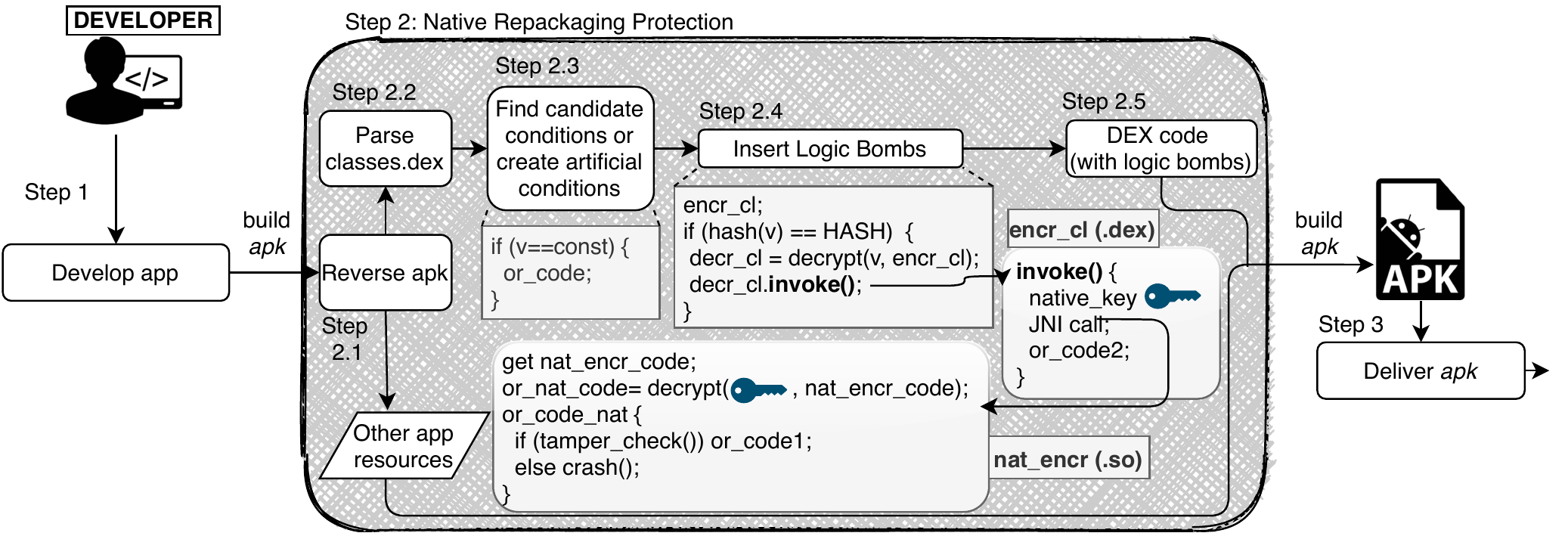}
  \caption{The Native Repackaging Protection scheme.}
  \label{fig:native-bombdroid-evolution}
\end{figure}

\paragraph{Protection scheme:} 
The NRP protection scheme, which mostly resembles the BombDroid and SDC ones, is depicted in Figure \ref{fig:native-bombdroid-evolution}.
Also in this case, NRP reverses the $apk$ (Step 2.1), parses the \longstring{classes.dex} file (Step 2.2), and finds (or builds artificial) blocks with a qualified condition. Then, it modifies each selected block as follows: i) given the standard condition \longstring{v==const}, NRP calculates the hash value of the constant (i.e., \longstring{HASH = H(const)}) using SHA-1 as the cryptographic hash function. Then, ii) the instructions of the block code (i.e., \longstring{or\_code}) are inserted in a new Java class which is encrypted using AES-128 in CTR mode, using the $const$ value as key. NRP applies several workarounds to preserve the semantic equivalence between the original code and the code contained in the encrypted Java class. Such class (i.e., \longstring{encr\_cl}) contains a single method \longstring{invoke()} which splits the original block code into two parts, namely \longstring{nat\_encr} and \longstring{or\_code2}. The first part refers to a native library which is likewise XOR-encrypted with a random symmetric key (i.e., $native\_key$). Such key is stored in the same encrypted Java class. To decrypt and execute the function \longstring{or\_nat\_code()}, stored in the \longstring{nat\_encr} library, a JNI call is required. \longstring{nat\_encr} contains an anti-tampering check that executes the first part of the original block code only if no tampering is detected. In current implementation, the integrity check verifies the signature of the first 100 bytes of the \longstring{classes.dex} file, whose path is hard-coded in the native library.

\paragraph{Runtime behavior:} During execution, the triggering of a logic bomb, depicted in Step 2.4 in Figure \ref{fig:native-bombdroid-evolution}, leads to the following execution: i) once the condition on the hash value is verified (i.e., \longstring{hash(v)==HASH}), the encrypted Java class (\longstring{encr\_cl}) is decrypted according to the constant value, and then the \longstring{invoke()} method is executed through Java reflection. The first statement (\longstring{nat\_encr}) is executed, and leads to the decryption of the native library (\longstring{nat\_encr}) and the execution of the \longstring{or\_code\_nat} block. Then, the signature of the first 100 bytes of the \longstring{classes.dex} file are checked. If the verification succeeds (i.e., \longstring{tamper\_check()==true}), then the first part of the original block code is executed (\longstring{or\_code1}), otherwise the app crashes. After the proper execution of \longstring{or\_code1}, the execution flow is returned to the \longstring{invoke()} method which executes the second part of the original block code (i.e., \longstring{or\_code2}).

\paragraph{Evaluation:}
The authors tested NRP against 100 apps downloaded from the Google Play Store and F-Droid. Tests were performed on an LG Nexus 5X device, running Android 8.1.0. In their implementation, authors leverage the \longstring{dalvik.system.InMemoryDexClassLoader} to load the new Java class of each encrypted block. This functionality is available starting from Android 8.0, only. The empirical assessment showed that 47\% of apps could be transformed without runtime exceptions. Each app has been tested through Monkey \citep{monkey} to generate random user's input: the time overhead is from 1.42\% up to 10.42\% (the space overhead is negligible), while on average, from 11 to 35 logic bombs have been activated, as shown in \cite{10.1007/978-3-030-45371-8_12}.

\section{Attacking Anti-Repackaging} \label{sec:attacking_model}
In this section, we discuss the \textsl{weapon rack} of repackaging, i.e., the main set of techniques that an attacker can leverage to analyze a protected $apk$ to detect and dismantle anti-repackaging protections (Steps 8 and 9 in Figure \ref{fig:threat-model-design}). It is worth pointing out that the same techniques can be likewise leveraged to detect malicious apps \citep{10.1007/978-3-030-01535-0_19}, as malware in Android often uses logic bombs to check proper environmental conditions before activating the payload (e.g., \cite{7546513}).

The analysis techniques can be divided into two main categories, namely \textsl{static} and \textsl{dynamic}. The first category refers to the set of techniques that analyze the $apk$ without executing it (static analysis), while the latter category is composed of techniques that apply to an $apk$ while executing on an actual or emulated Android device (dynamic analysis). 

On one hand, static analysis aims at systematically reverse engineering the $apk$ (e.g., White Box Reverse Engineering  ~\citep{mitre_white_box_reverse_engineering}) to reconstruct and analyze the original source code. There exist a lot of tools that support this type of analysis and help an attacker to decode, rebuild and inspect both the app code and the resources, such as apktool~\citep{apktool} and dex2jar~\citep{dex2jar}.
On the other hand, dynamic analysis aims at inferring the app behavior during execution, to retrieve, e.g., actual values of variables, and monitor the execution of the code. 

Static and dynamic analysis are complementary: the first aims at building a comprehensive model of the app behavior by systematically exploring all the code and the resources of the app; the latter focuses on inferring a model of the app behavior by repeatedly executing the app. In principle, the static model is a superset of all possible behaviors of the app, meaning that at runtime the app will exhibit only a subset of the behavior of the static model. The dynamic model is partial by definition (as it is impossible to stimulate all possible executions of the app at runtime) but contains only behaviors that the app exhibits. As a consequence, static analysis suffers from \textsl{false positives} (e.g., a statically detected behavior could not be executed at runtime), while dynamic analysis suffers from \textsl{false negatives} (a behavior which has not been recognized during the dynamic analysis phase could be assumed as non-existent). 
Static and dynamic techniques have been thoughtfully studied in recent years and further analysis is outside the scope of this paper. Nonetheless, the interested reader may refer to \citep{LI201767, GAJRANI202073, 10.1145/2996358} for more information.

Concerning anti-repackaging, static analysis aims to detect the pieces of code where logic bombs could potentially hide (e.g., blocks with hash values in the qualified condition), while dynamic analysis allows to check the actual triggering and execution of the bombs at runtime. Furthermore, the repeated execution of the app on several setups allows the attacker to carry out \textsl{cumulative attacks}~\citep{8368344}\footnote{See the AppIS approach in Section \ref{subsec:appis-2017}}.  
Within the ``try and error'' cycle, the attacker builds up an analysis workflow of the $apk$ which fruitfully combines static and dynamic analysis techniques.

Hereafter, we briefly introduce the main static and dynamic analysis techniques that are specifically tailored to detect and dismantle anti-repackaging controls.

\subsection{Code Analysis and Pattern Matching} \label{sec:text_search_and_pattern_matching}
This technique statically analyzes the decompiled $apk$ to detect keywords or expressions that may hide logic bombs and detection nodes. The search and detect phase relies on heuristics and regular expressions defined by the attacker herself, or on pre-defined patterns defined in static analysis tools that support this technique. For instance, this technique may allow recognizing explicit anti-tampering function calls, as well as qualified conditions containing hash values. As with any other static analysis approach, all detected anti-tampering checks must be evaluated at runtime to filter true positives (i.e., the checks which hide some logic bombs). 

The efficacy of this technique is rather limited, as it can be easily circumvented by anti-repackaging techniques. For example, SSN and BombDroid~\citep{7579771, 10.1145/3168820} are resilient to this threat as they rely on Java reflection or obfuscation~\citep{Norboev2018OnTR}; the first allows hiding function calls inside the code, while the latter avoids text search as obfuscation modifies the syntax of the code (e.g., method renaming).

\subsection{Dynamic Analysis \& Testing}\label{sec:fuzzing}
This category includes both automated techniques and manual testing. Attackers exploit dynamic analysis to test the app and exploit its behavior at runtime. We refer to ~\cite{8453877} for an overview of the state-of-the-art works in the Android ecosystem. 
Dynamic testing is suitable for assessing an $apk$ protected with SSN or Bombdroid, as it allows to trigger (and detect) the bombs (repackaging detection nodes) or perform analysis of hash values. However, these techniques do not grant a complete path exploration and full input data coverage. Therefore, they do not allow finding and triggering all bombs~\citep{Norboev2018OnTR}. Dynamic analysis allows resolving the recipient of a Java reflection call or getting access to dynamically loaded code~\citep{10.1145/2699026.2699105}. A generalization of dynamic analysis for software testing is known as \emph{Fuzzing}, which provides pseudo-random input to a running program.

\subsection{Debugging and Code Emulation}\label{sec:debugging_and_code_emulation}
The most common way to deeply analyze the behavior of an app is to use a debugger. In the general case, a debugger allows a developer to dynamically test software to find and resolve bugs. Concerning the attack to anti-repackaging, a debugger allows bypassing all static countermeasures like obfuscation, and understand how the app behaves at runtime.
More in detail, an attacker can intercept critical calls, find out repackaging detection, and understand the type of anti-tampering checks.
However, debugging strongly relies on manual intervention, which has the drawback to scale badly, i.e., debugging an app to try tracing back all bombs has been proven to be unfeasbile~\citep{Norboev2018OnTR, 10.1145/3168820}.

Besides, the debuggable flag inside the app manifest file has to be set to true to attach a debugger to an Android app. Developers that apply anti-repackaging techniques trivially set this flag to false. Therefore, an attacker should first modify the app manifest~\citep{owasp_mobile_security_testing_guide} and repackage it before debugging. This first repackaging step can itself trigger proper logic bombs in the app and make the execution of the app fail. 

An alternative to debugging is \textsl{code emulation} in \cite{180237, BHANDARI2017392}: the idea is that the target app is executed on some virtual machine, where the hardware, the Android OS, and the Dalvik VM layers are replicated. During app execution, the virtual machine performs call tracing \citep{10.1145/1966445.1966460}. 
However, both approaches follow specific execution paths that depend on input data and could not reach the candidate methods in which logic bombs are hidden.

\subsection{Process Exploration}\label{sec:process_exploration}
\textsl{Process Exploration} refers to a set of techniques allowing the attacker to deeply inspect the app process memory at runtime~\citep{owasp_mobile_security_testing_guide} to find out the detection nodes spread across the app. There exists several tool to support memory exploration, most of which are Frida-based\footnote{We will discuss Frida in Section  \ref{subsec:code_instrumentation_and_code_injection}}, and rely on \textsl{dynamic code instrumentation}.
Currently, the most widespread tools are:
\begin{itemize}
    \item \textbf{Fridump}: an open source memory dumping tool for both Android and iOS ~\citep{fridump}. From a memory dump, it is possible to access memory addresses where, for example, decryption key of anti-repackaging tools are stored at runtime;
    \item \textbf{Objection}: a runtime mobile exploration toolkit for both Android and iOS, that does not require a rooted device. Two interesting features of Objection are the ability to i) circumvent SSL pinning, i.e., a technique allowing apps to authenticate the server and avoid man-in-the-middle-attacks), and ii) dump and patch  memory locations~\citep{objection};
    \item \textbf{R2frida}: a tool that merges the reverse engineering capabilities of \textsl{radare2}~\citep{team2017radare2} (a reverse engineering framework) with Frida (a dynamic instrumentation toolkit). In this way, an attacker can carry out runtime reverse engineering and memory related tasks, such as memory inspection~\citep{r2frida}.
\end{itemize}
The adoption of such tools allows an attacker to analyze the runtime behavior of an app without debugging it (i.e., without the need to force the set of the debuggable flag to true and repackage.).

\subsection{Local OS tampering}\label{sec:vtable_hijacking}
Such techniques rely on features provided by the Android OS to detect and bypass some detection nodes. For instance, controlling the Java virtual table ($vtable$) provides several advantages to the attacker.

In Java, the vtable dynamically maps function calls, i.e., a Java object points to some records of the vtable which contains the link to the actual method implementation. In Java, all methods are virtual by default, unless they are declared as \textsl{final} or \textsl{static}.
An attacker can modify the values in the vtable to invoke an arbitrary method and change the behavior of the original method. 
The limitation of this approach is that it works properly only on rooted devices, as the attacker needs root privileges to change the entries of the vtable~\citep{artdroid}.

An example of vtable hijacking is proposed in~\cite{7579771}: here, the vtable mapping is modified to redirect all invocation to the \emph{getPublicKey} method to a spoofed function which returns the public key of the original developer, instead of the public key of the attacker that repackaged the app. 

\subsection{Code Manipulation}\label{subsec:code_manipulation}
There exist three main techniques to affect the app code in order to remove  anti-repackaging checks or to add malicious code, namely \textsl{code deletion}, \textsl{binary patching} and \textsl{code instrumentation and injection}. 

\paragraph{Code deletion:}\label{subsec:code_delection}
This technique aims at removing proper part of the app source code (if available), or smali code to deactivate specific controls. It can be partially circumvented: for instance, the anti-repackaging checks may be hosted in methods that cannot be straightforwardly removed, i.e., many other parts of the app depend on them. Alternatively, fake controls resembling logic bombs can be added in the code to fool the attacker which exploits code deletion \citep{10.1145/3168820}. In case the attacker removes them from the code, the app will not work properly.

\paragraph{Binary patching:}\label{paragraph:binary_patching}
It refers to the process of modifying a compiled executable to change its behavior. Different from code deletion, binary patching does not need to retrieve or reconstruct the app source code, and it is, for instance, independent from code obfuscation.
It is worth pointing out that it is far easier and more reliable to apply binary patching to Java than native code (e.g., C, C++, \ldots). Therefore, this technique is promising for an attacker, as current apps still contain most of the business logic in the Java/Kotlin part of the app (see Section \ref{sec:background}) rather than in the native part.

To counteract binary patching, the anti-repackaging checks need to be protected, i.e., through encryption applied at Java (see Sections \ref{subsec:dyn-self-prot-encrypt-2015}, \ref{subsec:sdc-2018},  \ref{subsec:bombdroid-2018}) or native level (Section \ref{subsec:bombdroid-native-2019}).
As an alternative, checks can be hidden by exploiting Java reflection~\citep{7579771}, however, once the attacker understands the protection pattern, he could patch the executable and remove checks with minimum effort, as discussed in~\cite{Norboev2018OnTR, 10.1145/3168820}.

\paragraph{Code Instrumentation and Injection:} \label{subsec:code_instrumentation_and_code_injection}
Such techniques allow to dynamically modify the code of a running process. In this way, the attacker can analyze functions or system calls, as well as their parameters and return values, and modify the behavior of the original functions. 

There exist a lot of tools that support code instrumentation and injection, as well as the corresponding runtime analysis. Nonetheless, Xposed \citep{xposed} and Frida \citep{frida}  are currently the \textsl{de-facto} standards.
Xposed is a framework that applies add-ons (called \textsl{modules}) directly to the Android OS ROM and requires root privileges. Such modules may allow, e.g., customizing the script that spawns any new process from the  Zygote one\footnote{The Zygote process is responsible for spawning new process to host launching apps.}, in order to add a jar file that allows hooking.

Frida is a dynamic code instrumentation tool that allows an attacker to hook functions by injecting a JavaScript engine into the instrumented process; the engine then allows injecting executable code by directly modifying the process memory. Frida can work on both rooted and unrooted devices in three different modes:
\begin{itemize}
    \item \textsl{Injected mode}: this is the most common deployment in which the user installs and runs the frida-server (i.e., a daemon) into a rooted device that exposes the frida-core module over TCP (usually on port 27042). 
    \item \textsl{Embedded mode}:  the user repackages the app with the frida-gadget (i.e., a shared library). This library allows instrumenting the app on an unrooted device.
    \item \textsl{Preloaded mode}: the frida-gadget shared library is injected inside the operating system layer instead of the app layer (as in the embedded mode). It works on rooted devices only.
\end{itemize}

Code instrumentation and injection allow stealing the actual values of program variables from a running app.
For example, the attacker could hook the function that decrypts some portion of code. In this way, he can recover the key for that specific chunk and the plain text. 
Then, the attacker can replace the encrypted code of an anti-repackaging check with a crafted one, by exploiting the binary patching technique previously discussed~\citep{owasp-mstg}.  

From the anti-repackaging standpoint, it is a hard task to safely hide checks that are resilient to code instrumentation and injection. As a matter of fact, system call hooking allows in principle to detect and attack any anti-repackaging check even if it is based on some kernel-provided functionality. Frida allows hooking tampering detection functions and makes them return the attacker's desired value. Also, some “anti-frida” controls (e.g. simply check if the TCP port 27042 is open) rely on system call and can be circumvented by Frida itself.

An attempt to limit function hooking is proposed in \cite{10.1007/978-3-030-45371-8_12}, where some Java instructions are translated into native code to elude Java code instrumentation and avoid an attacker to recover the original statements. 

\subsection{Network-based attacks}\label{subsec:network-attacks}
Previous attacking techniques focus on the device, the OS, and the app. However, it is also possible to attack anti-repackaging at network level. Such attacks can be passive, i.e., the attacker does not affect the network traffic (e.g., \textsl{sniffing attack}) or active, i.e., the attacker willingly affect and spoof the traffic, (e.g., \textsl{replay attack}). 

\paragraph{Sniffing attack.}\label{paragraph:sniffing_attack}
It refers to the interception and the analysis of data exchanged by two parties on a communication channel (e.g., an app interacting with a remote server)~\citep{capec-157, capec-609}. 
This attack is relevant as some anti-repackaging checks are executed on server side (i.e., external anti-tampering checks). If an attacker understands the communication pattern and which data are sent between parties, he could be able to detect and circumvent these controls~\citep{owasp-m3}. 
For instance, an attacker is able to read the client-server communication if it is over \textsl{http}.

\paragraph{Man in the Middle attack.}\label{paragraph:replay_attck}
The Man in the Middle attack extends the sniffing attack by actively interacting with the two parties.  The attack dynamically affects the data exchanged during the communication, by relying on a MitM proxy~\citep{capec-94}.
Concerning anti-repackaging, if the signature check of an $apk$ is carried out on server side, then the client has to send the checksum of its certificate and an attacker could change the communication data to trick the server or spoof a fictitious response from the server. Albeit all anti-repackaging techniques discussed in Section \ref{sec:sota} leverage local checks only, there exists some proprietary solutions which provide remote anti-repackaging detection, like Google SafetyNet~\citep{safetynet}. The safetyNet API is provided by Google to Android developers and can be included in any app. Such API provides functionalities against several security threats and specific checks against repackaging, which are applied both locally e.g., emulator detection, and remotely, e.g., signature detection on the server side. In this case, the attacker can try modifying the network traffic generated by the SafetyNet API calls, to bypass server-side checks or spoof fake answers from the server. Furthermore, the attacker can remove the invocation to SafetyNet API functions by directly patching the smali code (or native one) by leveraging code deletion (Section \ref{subsec:code_delection}).

It is worth pointing out that MitM attacks can be carried out also inside the $apk$ itself: in this scenario, the attacker could modify at runtime some data using some of the code/data manipulation techniques previously discussed, with the aim to bypass specific anti-repackaging checks. The main idea is that an attacker could replace some resource file in order to bypass specific controls which rely on the app data (e.g., modify the path from the repackaged resource into the original one)~\citep{Norboev2018OnTR, 8186215}.

\section{Disabling Anti-Repackaging} \label{sec:attack} 
\begin{table}[]
\begin{tabular}{|l|l|l|c|c|}
\hline
\multicolumn{1}{|c|}{\textbf{Protection scheme}} &
  \multicolumn{1}{c|}{\textbf{Year}} &
  \multicolumn{1}{c|}{\textbf{\begin{tabular}[c]{@{}c@{}}Detection and Neutralization\\techniques\end{tabular}}} &
  \multicolumn{1}{c|}{\textbf{\begin{tabular}[c]{@{}c@{}}Scheme\\bypassed\end{tabular}}} &
  \multicolumn{1}{c|}{\textbf{\begin{tabular}[c]{@{}c@{}}Scheme \\Implementation\\bypassed\end{tabular}}} \\ \hline
Self-Protection through dex Encryption &
  2015 &
  \begin{tabular}[c]{@{}l@{}}- Code Analysis and Pattern Matching \\ - Dynamic Analysis \& Testing\\ - Debugging and Code Emulation\\ - Process Exploration\\ - Code Manipulation\end{tabular} & 
  \begin{tabular}[c]{@{}c@{}}\checkmark\end{tabular} &
  \begin{tabular}[c]{@{}c@{}}\textcolor{black}{Implementation}\\\textcolor{black}{N/A}\end{tabular}  \\ \hline
SSN &
  2016 &
  \begin{tabular}[c]{@{}l@{}}- Code Analysis and Pattern Matching \\ - Dynamic Analysis \& Testing\\ - Debugging and Code Emulation\\ - Process Exploration\\ - Local OS tampering\\ - Code Manipulation\end{tabular} & 
  \begin{tabular}[c]{@{}c@{}}\checkmark\end{tabular} &
  \begin{tabular}[c]{@{}c@{}}\textcolor{black}{Implementation}\\\textcolor{black}{N/A}\end{tabular}  \\ \hline
AppIS &
  2017 &
  \begin{tabular}[c]{@{}l@{}}- Code Analysis and Pattern Matching \\ - Dynamic Analysis \& Testing\\ - Debugging and Code Emulation\\ - Process Exploration\\ - Code Manipulation\end{tabular} & 
  \begin{tabular}[c]{@{}c@{}}\checkmark\end{tabular} &
  \begin{tabular}[c]{@{}c@{}}\textcolor{black}{Implementation}\\\textcolor{black}{N/A}\end{tabular}  \\ \hline
SDC &
  2018 &
  \begin{tabular}[c]{@{}l@{}}- Code Analysis and Pattern Matching \\ - Dynamic Analysis \& Testing\\ - Debugging and Code Emulation\\ - Code Manipulation\end{tabular} & 
  \begin{tabular}[c]{@{}c@{}}\checkmark\end{tabular} &
  \begin{tabular}[c]{@{}c@{}}\textcolor{black}{Implementation}\\\textcolor{black}{N/A}\end{tabular}  \\ \hline
Bombdroid &
  2018 &
  \begin{tabular}[c]{@{}l@{}}- Code Analysis and Pattern Matching \\ - Dynamic Analysis \& Testing\\ - Code Manipulation\end{tabular} & 
  \begin{tabular}[c]{@{}c@{}}\checkmark\end{tabular} &
  \begin{tabular}[c]{@{}c@{}}\textcolor{black}{Implementation}\\\textcolor{black}{N/A}\end{tabular}  \\ \hline
NRP &
  2019 &
  \begin{tabular}[c]{@{}l@{}}- Code Analysis and Pattern Matching \\ - Dynamic Analysis \& Testing\\ - Debugging and Code Emulation\\ - Code Manipulation\end{tabular} & 
  \begin{tabular}[c]{@{}c@{}}\checkmark\end{tabular} &
  \begin{tabular}[c]{@{}c@{}}\checkmark\end{tabular} \\ \hline
\end{tabular}
\caption{Attacking vectors to anti-repackaging schemes.}
\label{table:protection_schema_attacks_recap}
\end{table}


In this section we discuss which set of the attacking techniques previously discussed can be leveraged to disable the anti-repackaging techniques presented in Section \ref{sec:sota}, and to build a fully working repackaged app.

Unfortunately, the vast majority of the older anti-repackaging techniques (i.e., described in Sections \ref{subsec:dyn-self-prot-encrypt-2015}, \ref{subsec:ssn-2016}, \ref{subsec:appis-2017}, \ref{subsec:sdc-2018} and \ref{subsec:bombdroid-2018}) have no available implementation to date. We tried to contact their authors via institutional emails, but we did not receive any positive answer (i.e., neither the source code nor a sample of a protected app). 
For this reason, we will just provide some motivating guidelines on how such techniques can be circumvented. 

However, the source code of the  native extension of BombDroid \citep{10.1007/978-3-030-45371-8_12}, described in Section \ref{subsec:bombdroid-native-2019}, is available as source code on GitHub \citep{repackaging-protection}. For this latter and more recent proposal, we describe a full-fledged attack able to circumvent all the anti-repackaging checks.

\textcolor{black}{In Table \ref{table:protection_schema_attacks_recap}, we summarized the outcomes of our security analysis. The table shows, for each anti-repackaging scheme, the corresponding publication year, which attacking techniques (discussed in Section \ref{sec:attacking_model}) can be used to detect and neutralize the implemented defenses (e.g., detection nodes), if we were able to develop a theoretical exploit (i.e., scheme bypassed column), and if we were able to execute that exploit on an available implementation of the schema (i.e., scheme implementation bypassed column).}

It is also worth noticing that none of the anti-repackaging techniques carries out any anti-debugging or emulation check. This basically simplifies the execution of cumulative attacks because the attacker can perform tests without worrying about emulator or debugging detection.

\subsection{Self-Protection through $dex$ Encryption}
The encryption method used to cipher the \texttt{classes.dex} file is a XOR encryption. This method could be easily extended to adopt more secure ciphers, like, e.g., AES; nonetheless, this would increase the runtime overhead.

Independently from the cipher, the main drawback of the proposal is that the  decryption key of the ciphered bytecode (i.e., the \texttt{classes.dex} file) is hardcoded in an obfuscated (via LLVM/OLLVM) native activity.  However, there is no need to statically de-obfuscate the native code to retrieve the key, as three dynamic analysis attacking techniques can be leveraged:
\begin{itemize}
    \item symbolic execution could reveal the decryption key and how it is obtained. This technique allows tracking several execution flows showing the value of the variables, including the decryption key;
    \item the debugger allows to detect the actual value of the key at runtime. To debug the app, the only modification is to change the value of the debuggable flag into the \emph{AndroidManifest.xml} file. This tampering is not detected by the XOR-based checksum at runtime;
    \item code instrumentation could allow retrieving the decryption key which is passed as a parameter to the decryption function. Function calls could be intercepted with different tools (e.g., see the Frida script in Figure \ref{fig:native-dump-argument-frida}) to report back the values of its arguments. 
    \begin{figure}[!htbp]
    \begin{lstlisting}[language=Java,
    basicstyle=\footnotesize\ttfamily,
    numbers=left,
    stepnumber=1,
    showstringspaces=false,
    tabsize=1,
    breaklines=true,
    breakatwhitespace=false,
    ]
    function inspectModule(module) {
      var m = Process.getModuleByName(module);

      Interceptor.attach(m.getExportByName("decrypt"), {}
        onEnter: function(args) {
          // Dump arguments
          send("Argument 1 of decrypt function" , args[0]);
          send("Argument 2 of decrypt function" , args[1]);
          [...]
        }
      });
    }
    \end{lstlisting}
    \caption{Frida script to dump arguments provided as input to the native method $decrypt$.}
    \label{fig:native-dump-argument-frida}
    \end{figure}
\end{itemize}
The access to the decryption key allows the attacker to decrypt any slice of ciphered code.
Furthermore, cumulative attacks can be performed to retrieve all the decrypted code using runtime memory dumping and process exploration. In this scenario, the attacker can dump each decrypted bytecode at runtime, thereby recovering the original code and she can leverage code manipulation (e.g., code deletion or binary patching) to add malicious code or remove anti-tampering controls. 

Once the code is recovered and tampered, one of the following approaches can be leveraged to create a working tampered $apk$ based on code deletion and binary patching:
\begin{itemize}
    \item override the body with the encrypted tampered code. This approach can be used only if a symmetric-key algorithm is used to encrypt and decrypt the code. In this way, once the attacker recovered the encryption key, she can perform both encryption and decryption.
    \item replace directly the body with the tampered code. In this scenario, the final app does not contain encrypted code.
\end{itemize}

\subsection{Stochastic Stealthy Networks} 
Several works \citep{Norboev2018OnTR, 10.1145/3168820, 10.1007/978-3-030-45371-8_12} have already pointed out some weaknesses of SSN, as well as some attacks that can be used to circumvent this protection. 
 There are two ways to circumvent SSN, namely i) deactivating the integrity check or ii) detect and remove all the detection nodes from the code.

Concerning the first strategy, recall that each detection node implements the same integrity check on any candidate method, based on the developer's public key. Therefore, it is sufficient to analyze a single detection node to define a de-activation strategy. To this aim, code injection or vtable hijacking allow replacing the target method (\textit{getPublicKey} and \textit{generateCertificate}) with a custom implementation, which returns the original developer's public key, instead of the attacker's one. 

Regarding the latter strategy, blackbox fuzzing can be used to generate a massive number of pseudo-random inputs to trigger all repackaging detection nodes (which are invoked stochastically) to detect them. In addition, whitebox fuzzing (or symbolic execution) can reveal the destination of the reflection calls.
Once all detection nodes are detected, either code deletion (i.e., to remove the repackaging detection nodes) or binary patching (i.e., to make the repackaging detection node unreachable) can be leveraged.

\subsection{AppIS: Protect Android Apps Against Runtime Repackaging Attacks}
The simplest way to repack an AppIS-protected app is to modify only parts (methods, resources, or data), which are not protected by guards. It is possible to recognize whether an app snippet is protected through static analysis, as no code obfuscation is applied by AppIS.
Static analysis also allows obtaining the set of guarding nets, and delete them from the code. This is made possible by the fact that each guard has a specific design pattern which the attacker can detect through bytecode analysis. In future works, the authors put forward the possibility to encrypt the set of guarding nets and decrypt it at runtime. Nonetheless, the encrypted guarding nets would be likewise vulnerable to deletion attack, as it is sufficient for the attacker to remove the parts of code where the decryption is carried out.

Beyond removing guarding nets, it is also possible to bypass the integrity detection on the goals, by exploiting cumulative attacks to collect a large number of the original hash values. Then, such values can be hardcoded in the corresponding integrity detection methods. In this way, the attacker can bypass the anti-tampering controls without deleting the guarding nodes. 

\subsection{Self-Defending Code through Information Asymmetry}
Recall that this approach is an improvement of the previous one: it moves the integrity checks to the native code and uses cryptographic obfuscation to resist static analysis.
Nonetheless, it remains vulnerable to several dynamic analysis techniques: first, blackbox fuzzing allows triggering a large number of SDC segments. For each triggered segment, a code instrumentation tool like Frida can be leveraged to hook the \texttt{encrypt} function in the \texttt{if} statement in order to recover the correct value of the variable \texttt{v}. 
 Once this data is obtained, it is possible to retrieve the decryption key of the encrypted code, in both modes supported by this anti-repackaging technique. Then, code deletion can be leveraged to substitute an SDC segment with the original code, thereby deactivating the protection.
Previous attacks succeed only if all SDC sufficient for repackaging succesfully are discovered.  
Therefore, the robustness of this anti-repackaging technique is directly dependent on the number and the distribution of SDC statements in the app.

Finally, if the attacker can locate the checksum method in the native code (the anti-tampering control), she would override the body of such method through binary patching, to force it returning the excepted value. In this way, we would be able to tamper an app and bypass all checks with low effort. 

\subsection{BombDroid: Decentralized Repackaging Detection through Logic Bombs} \label{subsec:attack-bombdroid}
In BombDroid, any logic bomb follows the same, well-defined, specific pattern (i.e., \longstring{hash(v)==const}). Text search and pattern matching can be used to statically analyze the app source according to this pattern to locate potential pieces of codes hiding logic bombs. However, code obfuscation could complicate this pattern-based search.
Once a potential logic bomb is detected, the attacker can try to retrieve the decryption key (which is the constant value of the qualified condition) through dynamic analysis by leveraging:
\begin{itemize}
    \item \textbf{Brute force}: The attacker can try all possible values for the integer value \texttt{v}. Potentially, this keyspace is huge (i.e., \texttt{$2^{32}$}, as Java stores integer values in 4 bytes). Nonetheless, the actual distribution of valid constant values is smaller \citep{repackaging-protection}. Furthermore, there is a small number of unique constant values (i.e., in the range [-1024, 1024], according to the authors). 
    \item \textbf{Extraction of information from open source code}: If an app includes external open source libraries, and the static analysis recognizes some potential logic bombs on such libraries, then an analysis of the original source code can allow guessing the right constant value of  \texttt{v}. 
    \item \textbf{Code instrumentation}: Proper code instrumentation applied on the potential logic bombs can allow intercepting the call to the decryption function (e.g., AES), thereby recovering the decryption key or the plain code (once the logic bomb has been triggered). 
\end{itemize}

Another way to retrieve the value of \texttt{v} is through cumulative attacks. In fact, as authors showed in \cite{10.1145/3168820}, most 9.3\% of bombs are triggered using both state of the art Android fuzzing tools and human analysis, the detection and de-activation of such few logic bombs can suffice to obtain a repackaged app that works properly in the most of the cases. Put another way, this fact indicates that albeit BombDroid can add a lot of logic bombs, only a few of them actually trigger.  Therefore, an attacker does not have to delete each logic bomb, but she needs to remove just the mostly triggered ones on average.

Once a constant value is retrieved from a single-trigger bomb, the logic bomb could be deleted (i.e., deletion attack) and replaced directly with the (original) decrypted code. The attacker can also add some extra code to the original one.
It is worth noticing that the previous attacking procedure applies to nested bombs if all constant values (i.e., contained in all nested qualified conditions) are discovered. 

Another possibility is to bypass the anti-tampering checks contained in the encrypted code of the bomb (i.e., the \longstring{tampering\_detection} function in Figure \ref{fig:bombdroid-scheme}):
\begin{itemize}
    \item leverage code injection to replace the tampering detection function with an attacker-defined one;
    \item try a replay attack to modify the MANIFEST.MF file used in the code digest comparison checks. In this scenario, the anti-tampering controls compare the hardcoded values with the ones in the original MANIFEST.MF file.
\end{itemize}

Recall that the BombDroid prototype relies on a tampering detection function based on a public-key comparison. Such function can be dynamically rewritten by instrumenting the Dalvik bytecode in order to return the expected value (i.e., the public-key of the original developer).

\subsection{Native Repackaging Protection}
\begin{figure}[!ht]
    \centering
    \includegraphics[width=\textwidth]{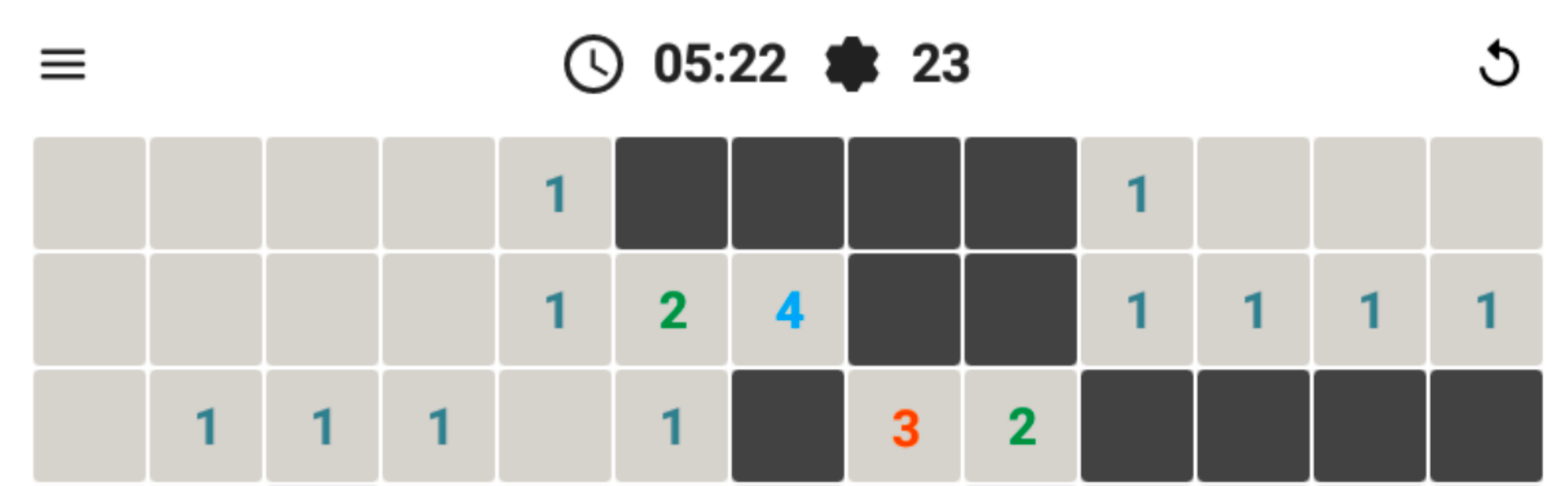}
    \caption{A snapshot of the Antimine game.}
    \label{fig:antimine-fig}
\end{figure}

The Native Repackaging Protection is the most recent and advanced repackaging avoidance technique to date, as it inherits and extends methodologies and techniques from the other proposals. 
For previous anti-repackaging techniques, we have pinpointed some high-level attack patterns by leveraging the attacking techniques described in Section \ref{sec:attacking_model}, since no source code, nor any protected app have been made available. 
However, as the source code of NRP is publicly available, we were able to analyze it in detail and carry out an actual and full-fledged attack to this technique.
We will describe the attack in details by relying on a real app, i.e., Antimine~\citep{antimine-android}, an open source, minesweeper-like game app available on F-Droid (see Figure \ref{fig:antimine-fig}).

More specifically, we will i) discuss some NRP implementation bugs, ii) apply NRP to Antimine, showing how a candidate method is protected, iii) prove how such protection can be circumvented and iv) build a full-working repackaged version of the protected Antimine app.

\subsubsection{NRP Implementation}
As a preliminary remark, we found an implementation bug by analyzing the NRP source code: the tool statically pre-computes the signature of the \longstring{classes.dex} file through a Java function, while at runtime the protected app calculates this value employing a native function (C/C++ code). The \texttt{char} type in standard C can either be signed [-128, +127] or unsigned [0, 255]. We found that the native function uses a char (line 3 on the left side of Figure \ref{fig:hash_bytes}) while the corresponding Java function adopts an unsigned char  (line 4 on the right side of Figure \ref{fig:hash_bytes}). As a consequence, a char value that has a bit representation with a \texttt{1*******} pattern (where \texttt{*} stands for 0 or 1) is interpreted as an int value between [-128,-1] by C and between [127,255] by Java. 
Therefore, if some chars of the \longstring{classes.dex} file have the \texttt{1*******} format, the integrity check always fails
- independently from the app being tampered or not. 

\begin{figure}
.
\begin{minipage}[t]{0.48\textwidth}
\begin{lstlisting}[language=C,
    basicstyle=\scriptsize\ttfamily,
    numbers=left,
    stepnumber=1,
    showstringspaces=false,
    tabsize=1,
    breaklines=true,
    breakatwhitespace=false,
    emphstyle=\underline,
    escapechar=ä,
]
// XOR in native code
int32_t hash_bytes(const char *arr, size_t count) {
  ä\underline{char hash[4];}ä// <-- Should be u_char
  for(size_t read = 0; read < count; read++) {
    hash[read % 4] ^= arr[read];
  }
  return hash[0] + (hash[1] << 8) + (hash[2] << 16) + (hash[3] << 24);
}
\end{lstlisting}
\end{minipage}
\hfill
\begin{minipage}[t]{.48\textwidth}

\begin{lstlisting}[language=Java,
    basicstyle=\scriptsize\ttfamily,
    numbers=left,
    stepnumber=1,
    showstringspaces=false,
    tabsize=1,
    breaklines=true,
    breakatwhitespace=false,
    emphstyle=\underline,
    escapechar=ä,
]
// XOR in Java code
private static int xorFile(Path p, int count, int offset) throws IOException {
  byte[] content = Files.readAllBytes(p);
  ä\underline{byte[] res = new byte[4];}ä
  for(int i = offset; i < (offset + count); i++ ) {
    res[(i - offset) % 4] ^= content[i];
  }
  return (0xff & res[0]) + ((0xff & res[1]) << 8) +
    ((0xff & res[2]) << 16) + ((0xff & res[3]) << 24);
}
\end{lstlisting}
\end{minipage}
\caption{On the left side: the C code used to calculate the hash sum at runtime. On the right side, the Java code used to precompute the hash sum, at compile-time.}
\label{fig:hash_bytes}
\end{figure}

Moreover, the translation of the Java bytecode into native code could introduce some new exceptions. For instance, any single instruction in the Java bytecode influence, or is influenced, by other instructions (for example the comparison between two variables). It is mandatory to manage this mutual relationship, and the program must behave soundly even if some instructions are executed under the native environment. We argue that this is a non-trivial and error-prone task, as no automatic tool currently exists to translate Java bytecode to native on Android. 
Furthermore, NRP suffers from the same limitations of BombDroid, related to the use of hash functions to hide the decryption key. If a logic bomb is inserted in an  \texttt{if} statement with a small hash space, it could be rather quick to recover the decryption key by brute force. As the authors discuss, NRP suffers from this limitation, as most of the constant values in an $apk$ are in the interval [-1024, 1024]. Finally, and differently from BombDroid, the hash function in NRP is not salted. As a consequence, whenever a decryption key is recovered, it is possible to statically detect all logic bombs that have the same key by only comparing the hash values. In fact, if two candidate conditions have the same constant value (e.g., \texttt{if(a == 1) {...}} and \texttt{if (b == 1){...}}, the hardcoded values in the trigger condition are likewise the same (i.e., \texttt{if(hash(a) == 'xyz')} and \texttt{if(hash(b) == 'xyz')},  since no salt is used. 

\subsubsection{Protecting Antimine with NRP}

We applied NRP to Antimine: the process leads to the insertion of $315$ logic bombs. For the sake of explanation, we focus on a single protected candidate method, namely the \texttt{setButton()} one. This method is invoked each time the user clicks on a blind button in the minesweeper game and changes the look and feel of the button and its neighbors, as well as the message to prompt to the user, according to the rule of the game (see Figure \ref{fig:antimine-fig}). As this method makes decisions, it contains several qualified conditions (e.g., \texttt{var == -1}). 

\begin{figure}
  \includegraphics[width=1.1\textwidth]{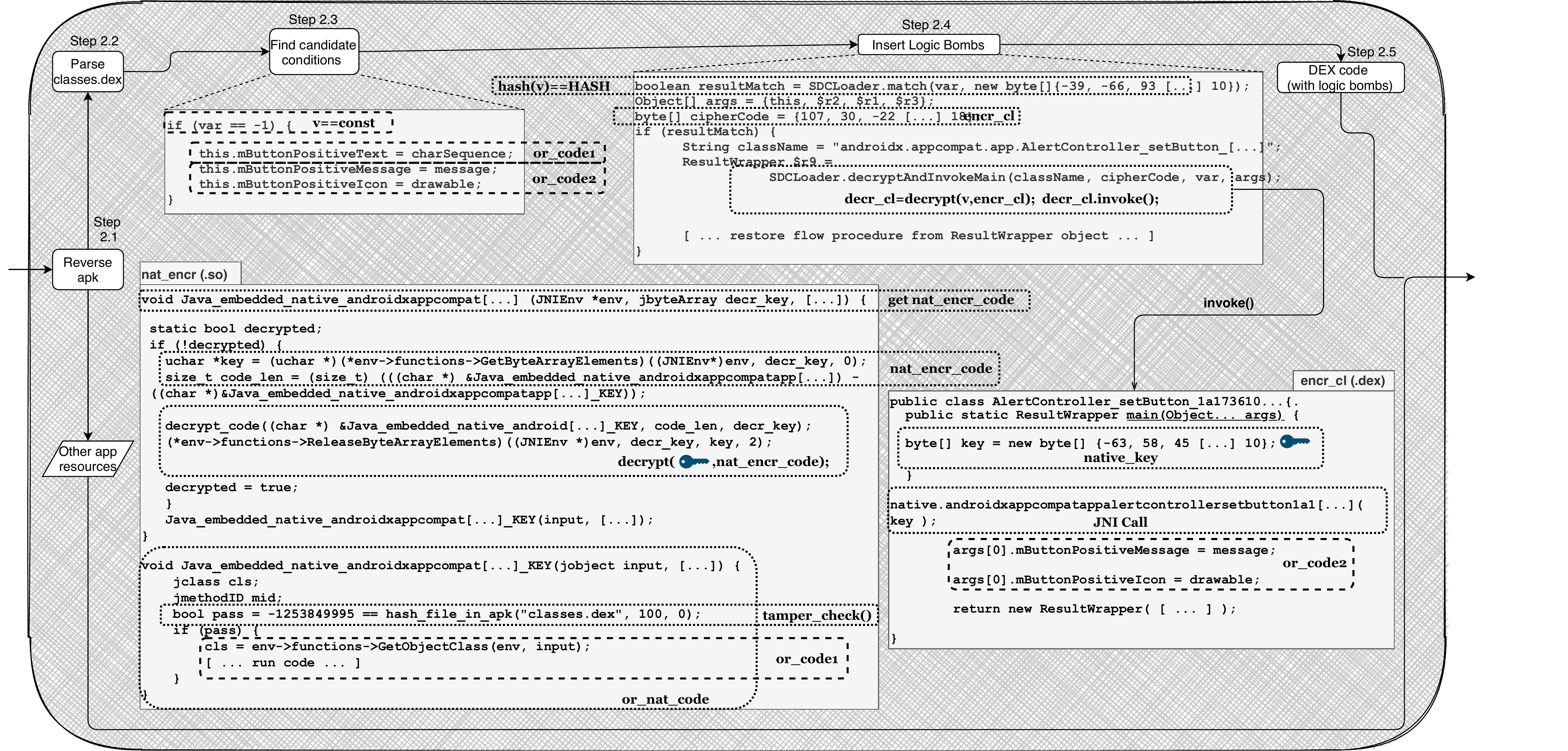}
  \caption{NRP applied to the \texttt{setButton()} method of Antimine.}
  \label{fig:native-bombdroid-evolution-example}
\end{figure}

Figure \ref{fig:native-bombdroid-evolution-example} describes the actual protection workflow, step-by-step. First, NRP parses the \texttt{classes.dex} file (Step 2.2) and selects all methods with at least a qualified condition as candidate methods (Step 2.3). NRP adds a logic bomb in any candidate methods and for each qualified condition. Concerning the \texttt{setButton()} method, it selects the condition \texttt{var == -1} and splits the original code of the corresponding block into two parts, namely \texttt{or\_code1} and \texttt{or\_code2}. 
In Step 2.4, NRP builds up the logic bomb by i) rewriting the if block in the candidate method, and ii) distributing the logic of the bomb among two encrypted files, namely a Java class (i.e., \texttt{encr\_cl.dex}) and a native C/C++ library (i.e., \texttt{nat\_encr.so}). Concerning the first point, the qualified condition is translated into a Boolean variable (i.e., \texttt{resultMatch}), whose value is calculated by comparing the \texttt{var} value with the hash of \texttt{-1} encoded as a byte array. In case the value of \texttt{resultMatch} is true, the block executes a \texttt{decryptAndInvokeMain(\ldots)} method. Such method takes as input the actual name of the encrypted Java class (i.e., \texttt{className = AlertController\_setButton...}), its binary representation (i.e., \texttt{cipherCode}, which is contained in the block), as well as the value of \texttt{var} and and the array of arguments (i.e., \texttt{args}) required by the main method of the encrypted Java class to execute properly. 
The invocation of such method leads to the decryption of \texttt{cipherCode}, using the value of \texttt{var} as decryption key, and to the execution of the main method of the class \texttt{AlertController...}. The main method contains the \texttt{native\_key} to decipher the native library (i.e., \texttt{nat\_encr.so}), a JNI call and the second part of the original block code (i.e., \texttt{or\_code2}). The JNI call triggers the execution of a deciphering function in the native library (i.e., \longstring{Java\_embedded\_native\_androidxappcompat(\ldots)}, which i) retrieves the encrypted native code from a memory location, ii) decrypts it using the \texttt{native\_key}, and iii) launches the execution of the decrypted function. This latter function (i.e., \longstring{Java\_embedded\_native\_androidxappcompat[\ldots]\_KEY(\ldots)}) contains the anti-tampering check (i.e., \texttt{hash\_file\_in\_apk(\ldots)}) that verifies the integrity of the first 100 bytes of the \texttt{classes.dex} file. If the integrity is verified, the function executes the first part of the original block code (i.e., \texttt{or\_code1}). Finally, the JNI call terminates and the main method in the Java class executes the last part of the original block (i.e., \texttt{or\_code2}).   

\subsubsection{Dismantling NRP bombs}
The attacker has to bypass the integrity checks performed by the protected app in the \longstring{Java\_embedded\_native\_androidxappcompat[\ldots]\_KEY(\ldots)} function to dismantle the NRP logic bombs.
To achieve such result, the attacker can either i) override \texttt{encr\_cl.dex} or \texttt{nat\_encr.so} for all the triggered logic bombs, by leveraging code deletion and binary patching, or ii) bypass the integrity check performed by the native code, by relying on a replay attack.

\paragraph{Bytecode and Native Code Overriding:}
The attacker can instrument the code to intercept the invocations to the \textsl{decryption functions} both in the Java and the native code, to retrieve i) the encrypted code and the decryption key passed as argument to the function, and ii) the decrypted code at the end of that function. Code instrumentation can also be used to change the implementation (or just the return value) of the tamper check function in the native code.

The Java decryption method requires two input values, namely the encrypted code (\texttt{param2}) and the decryption key (\texttt{param1}), and returns the decrypted code. 
Figure \ref{fig:linear-dump-frida} shows the Frida script that instruments the Java decryption method (i.e., \texttt{SDCloaderr.decryptAndInvoke(\ldots))} for a given encrypted code \texttt{encr\_cl}, represented by \texttt{ciphered\_byte} (i.e., the \texttt{cipherCode} byte array in the \texttt{setButton} method), that we will analyze using Jadx \citep{Jadx}). 
The script begins by selecting (lines 1 and 2) the class (i.e.,  \longstring{embedded.SDCLoader}) and the decryption method (i.e., \texttt{decrypt}) to overload. Frida hooks at the start of the overloaded function and executes the customized code (lines 3 to 19) instead of the original code.
The customized code calls the original Java \texttt{decrypt} method (line 6) and stores the returned value in a variable. Since the attacker aims to get access to a specific decrypted function, the script (lines 7-13) compares the encrypted method passed as the input parameter \texttt{param2} with \texttt{ciphered\_bytes}. If the decryption function is correct, the script writes down the decrypted bytecode (line 16). It is worth noticing that the script also dumps the key inside the \texttt{param1} variable. The return value of the overloaded function is the return value of the original decryption function (line 19).

\begin{figure}[!htbp]
\begin{lstlisting}[language=Java,
    basicstyle=\scriptsize\ttfamily,
    numbers=left,
    stepnumber=1,
    showstringspaces=false,
    tabsize=1,
    breaklines=true,
    breakatwhitespace=false,
]
var SDCLoader = Java.use("embedded.SDCLoader");
SDCLoader.decrypt.overload("[B", "[B").implementation = function(param1, param2) {
  var arr_bytes = Java.array('byte', param2);
  var ciphered_bytes = [107, 30, -22 [..] 18];
  var found = true;
  var dexBytesDecrypted = this.decrypt(param1, param2);
  if(arr_bytes.length == ciphered_bytes.length) {
  	for(var i = 0; i < ciphered_bytes.length; i++) {
  	  if(ciphered_bytes[i] != arr_bytes[i]){
  	    found = false;
  		break;
  	  }
  	}
    if (found) {
  	  console.log("### FOUND");
  	  // Write to file dexBytesDecrypted
    }
  }
  return dexBytesDecrypted;
}
\end{lstlisting}
\caption{Frida script to dump the result of decrypted function.}
\label{fig:linear-dump-frida}
\end{figure}

Concerning the native decryption function, it can be reversed using a decompiler (i.e., Ghidra \citep{ghidra}), and analyzed to detect the decryption function, and to infer the decryption logic (e.g., it is possible to analyze the body of the \texttt{Java\_embedded\_native\_androidxappcompat[...]} function). 
Here, the \texttt{decrypt\_code} function is responsible for decrypting $n$ bytes of machine code starting from a pointer to the memory. Such function can be overloaded and dumped through the scripts described in Figure \ref{fig:native-dump-frida} and Figure \ref{fig:native-dump-frida2}, which allow retrieving both the plain code and the key, as in the case of the Java decryption function. 
In detail, the script in Figure \ref{fig:native-dump-frida} aims to detect the location in which the target function to dump will be decrypted and executed. It starts by retrieving the \texttt{libnative-lib.so} (line 2) - which is the shared object containing all native functions implemented by NRP and then it enumerates each function therein (lines 6-10) to detect the \texttt{decrypt\_code} functions. Then, the script sets up an interceptor for the function (line 13). Whenever a \texttt{decrypt\_code} function executes, the script verifies whether the decryption function is the expected one (lines 14-30), by comparing the key passed as a parameter to the \texttt{native\_key} hardcoded in the \texttt{encr\_cl.dex} file. 
If this is the case, the script invokes the \texttt{AES\_CTR\_xcrypt\_buffer} function that decrypts the code, write the plain machine code in a writable page, and makes it executable. More specifically, the script in Figure \ref{fig:native-dump-frida2} attaches Frida to the \texttt{AES\_CTR\_xcrypt\_buffer} function (lines 5-14). Then, when the function starts (i.e., OnEnter, lines 15-18), the script retrieves the memory address (i.e., \texttt{args[1]}) and the size (i.e., \texttt{args[2]}) of the write buffer for the code. After the decryption phase terminates (i.e., OnLeave (from line 19 to line 24), the script checks whether the decrypted code is correct (line 20): in case, the script retrieves the pointer to the starting address (line 21) and dumps \texttt{len} bytes (lines 22-24).

\begin{figure}[!htbp]
\begin{lstlisting}[language=Java,
    basicstyle=\scriptsize\ttfamily,
    numbers=left,
    stepnumber=1,
    showstringspaces=false,
    tabsize=1,
    breaklines=true,
    breakatwhitespace=false,
]
function inspectModule(module) {
  var m = Process.getModuleByName(module);
  var targetFunction = true;
  var functionToBeIntercepted = "";
  var exports = m.enumerateExports();
  exports.forEach(function(exp){
    if (exp.name.indexOf("decrypt_code")>=0){
      functionToBeIntercepted = exp.name; 
    }
  });
  // Intercept decrypt_code function
  Interceptor.attach(m.getExportByName(functionToBeIntercepted), {
    onEnter: function(args) {
      var pointer = new NativePointer(args[2]);
      var buff = pointer.readByteArray(16);
      buff = new Int8Array(buff);
      var key = [-63, [...], -9]; // native_key hardcoded in encr_cl.dex
      targetFunction = true;
      for(var i = 0; i < buff.length; ++i){
        if(buff[i] != key[i]){
          targetFunction = false;
          break;
        } 
      }
      if(targetFunction) {
	    console.log("##### FOUND");
      } 
    }
  });
  // Intercept AES_CTR_xcrypt_buffer function
  [...]
}
\end{lstlisting}
\caption{Frida script that searches for the target \texttt{decrypt\_code} function.}
\label{fig:native-dump-frida}
\end{figure}

\begin{figure}[!htbp]
\begin{lstlisting}[language=Java,
    basicstyle=\scriptsize\ttfamily,
    numbers=left,
    stepnumber=1,
    showstringspaces=false,
    tabsize=1,
    breaklines=true,
    breakatwhitespace=false,
]
function inspectModule(module) {
  [...]
  // Intercept AES_CTR_xcrypt_buffer function
  // Dump the return bytes
  functionToBeIntercepted = "";
  exports = m.enumerateExports();
  exports.forEach(function(exp){
    if (exp.name.indexOf("AES_CTR_xcrypt_buffer")>=0){
      functionToBeIntercepted = exp.name; 
    }
  });
  var outAddr = "";
  var len = 0;
  Interceptor.attach(m.getExportByName(functionToBeIntercepted), {}
    onEnter: function(args) {
      outAddr= args[1];
      len = parseInt(args[2]);
    },
    onLeave: function(retval) {
      if(targetFunction) {
        var pointer = new NativePointer(outAddr);
        var buff = pointer.readByteArray(len);
        // DUMP buff
        send("Readed byte", buff);
      }
    }
  });
}
\end{lstlisting}
\caption{Frida script that dumps the assembly code decrypted by the native function.}
\label{fig:native-dump-frida2}
\end{figure}

\newpage

The decrypted code is a hex string that encodes the x86 assembly instructions. Such instructions can be extracted using a disassembler. 
The analysis of the disassembled instructions allows detecting a hardcoded tamper check on the \texttt{classes.dex} file 
\begin{center}
    \texttt{cmp eax, 0xb543c475}
\end{center}
where \texttt{$0xb543c475$} is the checksum of the original \texttt{classes.dex} file of Antimine.

Once the original codes and the decryption keys are recovered, an attacker can easily bypass the tamper check by overriding the hardcoded value with one of the repackaged app, and then re-encrypting the assembly code and substituting the original code in the $.so$ file. As an alternative, the attacker can modify the decrypted \texttt{nat\_enc.so} and the \texttt{encr\_cl.dex} to execute \texttt{op\_code1} in Java rather than in the native code, and remove the JNI call that would trigger the tamper check.  

\begin{figure}[!htbp]
  \includegraphics[width=\textwidth]{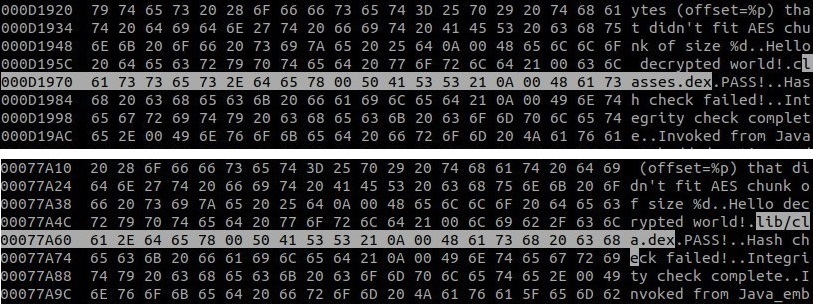}
  \caption{The original (top side) and the tampered (bottom side) library file.}
  \label{fig:sochanged}
\end{figure}

\paragraph{Bypass the integrity check:}
An alternative attack to bypass the tamper check is to leverage a code manipulation attack to perform a MitM-like attack by modifying the file on which the tamper check is executed. This attack can be carried out since the path of the checked file is hardcoded in the native library in cleartext. 

We carried out this attack as follows: we added a new file (i.e., \texttt{cla.dex}) to the $apk$, that has the same content of the original \texttt{classes.dex} of Antimine. Then, we manually patched the hardcoded path in the \texttt{elf} file to point to \texttt{cla.dex} instead of the \texttt{classes.dex} which is modified by the attacker before repackaging. In this way, the tamper check passes as it calculates the expected value, despite the app content has been modified. Figure \ref{fig:sochanged} shows the original library (top side), and the tampered one (bottom side) in which the hardcoded path is modified to point to $cla.dex$.

Previous attacks allowed to successfully bypass the NRP protection on Antimine. The repackaged version of Antimine, with all the 315 bombs de-activated, is available at:
\begin{center}
\url{https://www.csec.it/projects/antimine-800011-protected-and-repacked.apk} \\
\end{center}
for further studies, analysis, and testing activities. 

\section{Discussion, Conclusion and Future Work}
\label{sec:conclusion}

In this paper, we discussed the state of the art of anti-repackaging techniques, unveiling their limitations, and the way to attack them.
Such techniques are far from being robust against the available attacking vectors.

To make things worse, the Android ecosystem is prone to the spreading of repackaged apps since it is possible to retrieve apps from third-party - untrusted - sources. Furthermore, The weak app signing rules adopted by Google do not prevent repacked apps from being spread through the Google Play Store.

Possible mitigation to the repackaging threat could be introducing a centralized authority that is in charge of signing and verifying the trustworthiness of the delivered apps, as it happens in iOS environments. Unfortunately, such a paradigm appears to be in contrast with both the Google policies and business model and could discourage developers.

To this aim, the improvement of self-protecting anti-repackaging schemes on Android apps is of paramount importance to limit the spread of repackaged apps.
To this aim, we argue that improving the reliability of anti-repackaging on Android demands developing novel methodologies that can discourage an attacker from trying to deactivate them. Next-generation anti-repackaging approaches should follow some guidelines, derived from the lesson learned:
\begin{enumerate}
    \item \textbf{Use multiple patterns}. Current techniques adhere to a common pattern (i.e., qualified conditions and adoption of hash functions), which we showed to be rather easy to detect and de-activate by an attacker. The same anti-repackaging scheme should implement several distinct patterns to disseminate logic bombs in the app. 
    Furthermore, as we have shown in NRP, a single anti-tampering control could be easily bypassed by an attacker. Therefore, several heterogeneous anti-tampering controls must be distributed throughout the app.
    \item \textbf{Rely on native code}. As reversing and instrumenting native code is more complicated than bytecode, both logic bombs and their triggering conditions should be moved to native as much as possible. 
    \item \textbf{Optimized pervasiveness of the logic bombs}. Some approaches implement few controls in the name of performance, while others embed a lot of bombs that are not actually triggered at runtime. In this respect, there is a need for more effective deployment of logic bombs, also at the cost of higher time and space overheads.
    \item \textbf{Honeypot bombs}. Proper fake bombs can be added to disguise the attacker to remove them. The removal of such bombs should trigger the real ones and lead the app to crash. This idea has been proposed in literature but never actually applied in current proposals, where the focus is more on generating stealthy logic bombs. 
    \item \textbf{Hiding anti-repackaging protection}. It is fundamental to improve the stealthiness of anti-repackaging schemes to make it hard for the attacker to understand that the reverse-engineered app contains some protection. Therefore, obfuscating anti-repackaging and anti-tampering controls may help.
    \item \textbf{Mixed Client-side and Server-side controls.} Most of the proposals check the app directly on the device (client-side) while others adopt remote controls (server-side) only. A more effective anti-repackaging protection must combine both of them. Furthermore, runtime-generated server-side controls can improve undetectability.
    
    In contrast with client-side controls, the server-side ones are based on a secure communication channel and a trusted data source (i.e., the server relies on clients' data). As previously pointed out, an anti-repackaging approach exploiting only server-side controls can be bypassed by tampering with the communication channel. Nevertheless, server-side controls can be useful to create a more heterogeneous set of detection nodes and add code splitting. In this way, the attacker should adopt a different technique for each type of control, thereby requiring a higher amount of time to counteract anti-repackaging.
\end{enumerate}

As a future extension of this work, we plan to implement an anti-repackaging solution that we already designed according to the previous guidelines. We plan to assess it on a high number of Android apps and to make both the source code and the experimental results available to the research community.






\bibliography{main}

\newpage

\bio{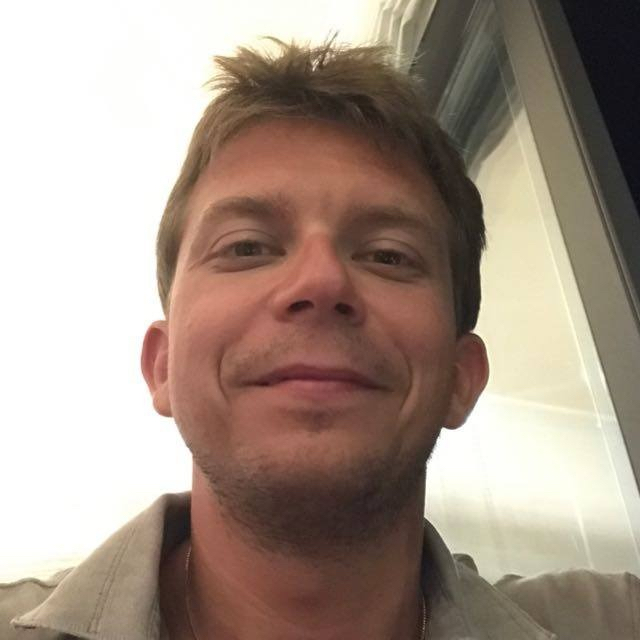}
\textbf{Alessio Merlo} is an Associate Professor at Computer Engineering in the Departiment of Informatics, Bioengineering, Robotics and System Engineering Department (DIBRIS) at the University of Genoa, and a member of the Computer Security Laboratory (CSEC Lab). His main research field is Mobile Security, with a specific interest on Android security, automated static and dynamic analysis of Android apps, mobile authentication and mobile malware. More information can be found at: \url{http://csec.it/people/alessio\_merlo/} 
\endbio

\vspace{0.2in}

\bio{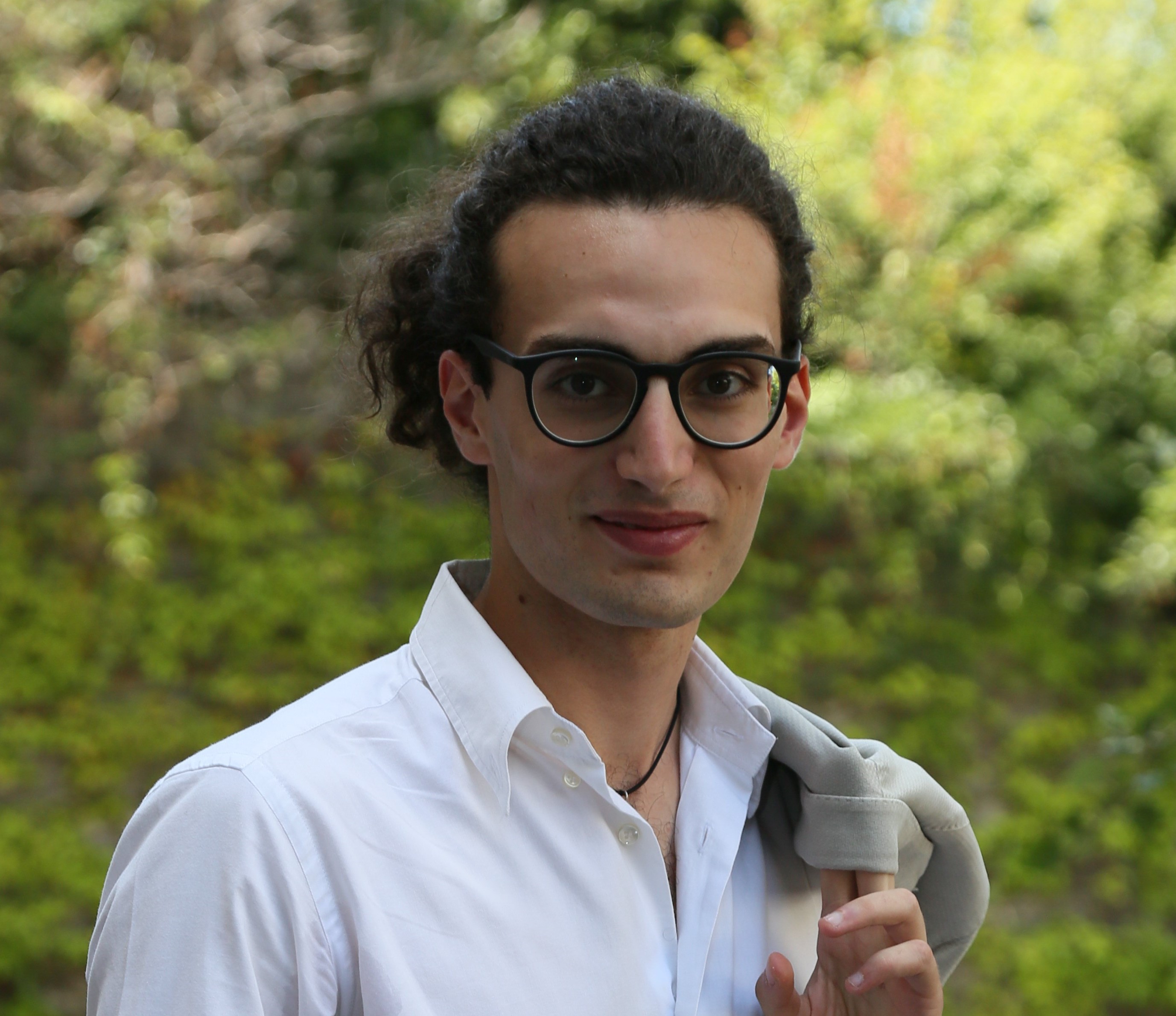}
\textbf{Antonio Ruggia} is a PhD student in Computer Engineering at the University of Genoa. He is interested in several security topics that include Mobile Security, with a specific interest in Android and data protection. He graduated in October 2020 at University of Genoa and participated in the 2019 CyberChallenge.it, an Italian practical competition for students in Cybersecurity. He regularly participates in international Capture-The-Flag (CTF) competitions. Since 2018, he has been working as a full-stack developer in a multinational corporation.
\endbio

\vspace{0.1in}

\bio{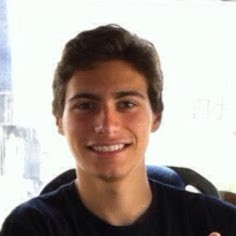}
\textbf{Luigi Sciolla} holds a Master degree in Computer Engineering at University of Genoa. During his studies he worked as a system administrator and full stack mobile developer. His area of interests are computer networks and Android security. He participated in the 2018 CyberChallenge.IT finals. Since 2018, he is also member of ZenHack, the capture-the-flag team at the University of Genoa. With ZenHack he competed in dozens of online and onsite competitions. 
\endbio

\vspace{0.25in}

\bio{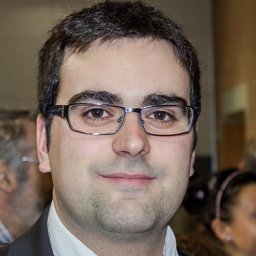}
\textbf{Luca Verderame} obtained a Ph.D. in Electronic, Information, Robotics, and Telecommunication Engineering at the University of Genoa (Italy) in 2016, where he worked on mobile security. He is currently working as a post-doc research fellow at the Computer Security Laboratory (CSEC Lab), and he is also the CEO and Co-founder of Talos, a cybersecurity startup and university spin-off. His research interests mainly cover information security applied, in particular, to mobile and IoT environments.
\endbio

\end{document}